\documentclass[%
 aip,
 rsi,
amsmath,amssymb,
reprint,
floatfix
]{revtex4-1}

\usepackage{graphicx}
\usepackage{dcolumn}
\usepackage{bm}

\usepackage[utf8]{inputenc}
\usepackage{mathptmx}
\usepackage{braket}
\usepackage[english]{babel}
\usepackage{xcolor}
\colorlet{RED}{red}
\colorlet{BLUE}{blue}
\usepackage{dcolumn}
\usepackage{bm}
\usepackage[version=4]{mhchem} 
\usepackage{acronym}
\usepackage{adjustbox}
\usepackage{tikz}
\usepackage[toc,page]{appendix}
\usepackage{microtype} 
\usetikzlibrary{calc,shapes.geometric,decorations.pathmorphing,patterns}

\definecolor{background-color}{gray}{0.98}
\usepackage[margin=2.3cm,bmargin=1cm,footnotesep=1cm]{geometry}

\newcommand{\PNNL}{%
   \affiliation{%
       Physical Sciences Division, Pacific Northwest National Laboratory, Richland, WA 99354, USA
       }}

\begin{document}

\title{
%
Dimensionality reduction of many-body problem using coupled-cluster sub-system flow equations: classical and quantum computing perspective
}

\author{Karol Kowalski} 
\email{karol.kowalski@pnnl.gov} \PNNL 

\date{\today}

\begin{abstract}
We discuss reduced-scaling strategies employing recently introduced sub-system embedding sub-algebras coupled-cluster formalism (SES-CC) to describe quantum many-body systems. 
These strategies utilize properties of the SES-CC formulations where the equations describing certain classes of sub-systems can be integrated into a computational flows composed of 
coupled eigenvalue problems of reduced dimensionality.    Additionally, these flows can be determined at the level of the CC Ansatz  by the inclusion of  selected classes of cluster amplitudes, which 
define the wave function "memory"   of possible partitionings of the many-body system into constituent sub-systems. 
One of the possible ways of solving these coupled problems is through
implementing  procedures, where the information is passed between the sub-systems in a self-consistent
manner. As a special case, we consider local flow formulations  where the 
local character of correlation effects can be closely related to properties of sub-system embedding sub-algebras employing localized molecular basis. 
We also generalize flow equations to the time domain and to downfolding methods utilizing double exponential unitary CC Ansatz (DUCC), where reduced
dimensionality of constituent sub-problems offer a possibility of efficient utilization of limited quantum resources in
modeling realistic systems.
\end{abstract}

\maketitle

%
%
%
%

\section{ Introduction} 
Over the last few decades, the coupled-cluster (CC) theory \cite{coester58_421,coester60_477,cizek66_4256,paldus72_50,purvis82_1910,jorgensen90_3333,paldus07,crawford2000introduction,bartlett_rmp} has 
evolved into one of the most accurate and dominant  theory to describe  various  quantum  systems across spatial scales
hence addressing fundamental problems in many-body physics \cite{arponen83_311,arponen1987extended1,arponen1987extended,arponen1991independent,arponen1993independent,arponen1993independent,
robinson1989extended,arponen1991holomorphic,bishop1978electron,bishop1982electron,arponen1988extended,emrich1984electron} (for an excellent review of these developments see Ref.\cite{bishop1991overview}),
quantum field theory,\cite{funke1987approaching,kummel2001post,hasberg1986coupled,bishop2006towards,ligterink1998coupled} quantum hydrodynamics,\cite{arponen1988towards,bishop1989quantum}
nuclear structure theory,\cite{PhysRevC.69.054320,PhysRevLett.92.132501,PhysRevLett.101.092502} quantum chemistry,\cite{scheiner1987analytic,sinnokrot2002estimates,slipchenko2002singlet,tajti2004heat,crawford2006ab,parkhill2009perfect,riplinger2013efficient,yuwono2020quantum} and material sciences.\cite{stoll1992correlation,hirata2004coupled,katagiri2005equation,booth2013towards,degroote2016polynomial,mcclain2017gaussian,wang2020excitons,PhysRevX.10.041043}  
Many  strengths of the single-reference CC formalism (SR-CC)  or coupled-cluster methods(CCM) originates in the exponential parametrization of the ground-state wave function $|\Psi\rangle$
\begin{equation}
|\Psi\rangle = e^T |\Phi\rangle \;,
\label{ccwf}
\end{equation}
where $T$ and $|\Phi\rangle$ correspond to cluster operator and reference function. For example, one can define a hierarchy of CC approximations by increasing the rank of excitations  included  the cluster operator. Another important feature of CC stems from the linked cluster theorem \cite{brandow67_771,lindgren12} which allows one to build efficient non-iterative  algorithms for higher-order excitations. 
When both  approximation  techniques are combined, 
one can define 
efficient and accurate methodologies that can deliver a high-level of accuracy in chemical simulations.
\cite{raghavachari89_479,stan1,stan2,crawford_t,gwal1,gwal2,gwal3,sohir1,ybom1,robkn,ugur1,mauss1,mebar1,mmcc1,crccx,bauman1}
More  recently, CC methodologies have been integrated with stochastic Monte Carlo methods probing configurational space and leading to near full-configuration-interaction accuracy of calculated energies.
\cite{thom2010stochastic,deustua2017converging,yuwono2020quantum}
However, the applicability of canonical formulations of these theories (especially to large molecular systems)
may be  limited by their steep (polynomial) numerical scaling. Unfortunately, even with rapid progress in computational technologies, problems with data locality, data movement,  and 
 polynomial scaling of high-rank canonical CC methods lead to insurmountable numerical problems in modeling large systems.
Although impressive progress has been achieved in the development of local approaches for CC pair theories,
\cite{hampel1996local,schutz2000low,schutz2000local,schutz2001low,neese2009efficient,neese2009efficient2,riplinger2013natural,Neese16_024109,pavosevic2016,pavosevic2017,saitow2017new,guo2018communication}
extension of these methods to include higher-rank excitations  may still require a significant  theoretical effort.
Some of these problems may be addressed by using the mature form of quantum computers, however, due to the limited size of existing quantum registers this can only be achieved by developing flexible algorithms 
to reduce the dimensionality of quantum problem. 
These problems have been scrutinized only recently, including local and reduced-dimensionality quantum computing formulations.
\cite{bauman2019downfolding,downfolding2020t,bauman2019quantumex,metcalf2020resource,bauman2020variational,takeshita2020increasing,motta2020quantum,mcardle2020improving,csahinouglu2020hamiltonian}

In the light of the above discussion, new high-accuracy  CC-based techniques for re-representing quantum many-body problem in reduced-dimensionality spaces are in high demand.
Especially interesting are  approaches where the original high-dimensionality problem can be recast in the form of  coupled low-dimensionality problems.
Also,  for quantum computing algorithms, the dimensions of  sub-problems coupled into a flow
should  be tunable to the available quantum computing (QC) resources to provide,   by controlling
the number of parameters processed at a given time, optimal utilization of computational tools such as  Variational Quantum Eigensolvers (VQE).
\cite{peruzzo2014variational,mcclean2016theory,romero2018strategies,PhysRevA.95.020501,Kandala2017,kandala2018extending,PhysRevX.8.011021,huggins2020non,cao2019quantum,ryabinkin2020iterative}
Additionally, recent strides made in the development of unitary CC formulations \cite{hoffmann1988unitary,unitary1,unitary2,harsha2018difference,lee2018generalized,izmaylov2020order,grimsley2019trotterized}
such as their disentangled,\cite{evangelista2019exact}  adaptive VQE variants,\cite{grimsley2019adaptive} 
and qubit representations of unitary CC methods and exact quadratic truncation of the Baker–Campbell–Hausdorff expansion \cite{ryabinkin2018qubit,lang2020unitary}
provide tools not only for next-generation VQE-type solvers but also for unlocking properties of unitary CC formulations needed in the analysis of reduced-dimensionality methods \cite{bauman2019downfolding,downfolding2020t}.



In this paper, we will focus on the further extension of  recently introduced CC sub-system embedding sub-algebras CC (SES-CC) \cite{safkk} and  double unitary CC downfolding methods (DUCC).\cite{bauman2019downfolding,downfolding2020t}
In a natural way, these methods allow to calculate ground-state energies as  eigenvalues of effective Hamiltonians in pre-defined  active spaces describing  sub-systems of the whole quantum system. 
Since  in the construction of   effective Hamiltonians  all out-of-active-space correlation effects are integrated out, the CC downfolding procedures can be viewed as a natural renormalization techniques. 
The flow equations for single-reference SES-CC case,\cite{safkk}  utilize this property for each active space involved in the flow
and for this reason  can be considered as formulations that have  "memory" of each  sub-system involved in the flow.  
Specifically, each sub-system can be characterized by the corresponding effective Hamiltonian that  includes interactions with other sub-systems.
We will illustrate  the ability of these approaches 
to capture complicated correlation effect and dynamics of the system, through traversing large sub-spaces of the entire Hilbert space without an unnecessary increase of the size of the  numerical problem to be solved at a given time in the flow algorithm.  We will also show that it is possible to define flows that decouple  the representation of the Schr\"odinger equation in large sub-spaces of the Hilbert (often defined by net dimensions beyond classical/quantum computing capabilities) into smaller problems that are numerically tractable. 


This paper discusses extensions of the CC flow formalism that have not been analyzed in earlier articles on this subject. Specifically, it includes:
\begin{itemize}
\item An explicit proof of the fact that the arbitrary  CC flow equations  are equivalent to standard CC equations with a specific form of cluster operator.
\item The extension of the CC flow algorithms to the time domain. In analogy to the previous point, we also show that time-dependent CC flow equations are equivalent to the time-dependent CC equations with a specific form of the cluster operator.
\item The extension of the CC flow equations to the localized basis set. In this case, we demonstrate that the CC flow equations provide a rigorous definition (at the level of effective Hamiltonians)  of the so-called electron pair. This type of CC flow in a natural way defines density matrices and higher-rank excitations for local CC formulations. The introduction of higher-rank excitations is a well-known problem faced by local formulations of CC methods. 
\item The extension of CC flows to double unitary CC (DUCC) formulations. We discuss these flows from the point of view of quantum computing applications, where DUCC flows can be used to probe configurational spaces of the dimensions beyond dimensions treated by existing quantum algorithms. Additionally, DUCC flow for localized orbitals naturally addresses Hamiltonian qubit encoding and controlling anti-symmetry of the corresponding wave function. 
\end{itemize}
Since the operator algebra involved in the unitary CC methods is non-commutative, extending canonical SES-CC flows to the DUCC case requires the utilization of certain approximations. 
To this end,  we will consider  backward-type methods  based on the use of approximate Trotter formulas.

We will also discuss the difference between two computational strategies involving  (1) standard approximations based on the selection of cluster amplitudes and treating them simultaneously (or globally) in numerical implementations and (2) flow equations where only a portion of selected amplitudes are processed at the time. While the former computing approach can take advantage of parallel classical architectures, the latter is ideally suited for  Noisy Intermediate-Scale Quantum devices (NISQ), where a small subset of fermionic degrees of freedom can be effectively handled. 
The flow equation methods also provide a conceptual foundation
for introducing certain approximations classes and eliminating possible problems with their postulatory character. We will illustrate these advantages on the example of local CC methods. 
For simplicity, in this paper we will focus on the CC and DUCC flow equations for closed-shell systems. 
\\

\section{\em Sub-system embedding sub-algebras CC  formalism - stationary and time-dependent formulations}

The SES CC formalism is based on the observation that energy of CC formulations, 
$E_{\rm CC}$, in addition to the well know formula $\langle\Phi|e^{-T}He^{T}|\Phi\rangle$ (where $H$ represents many-body Hamiltonian),   
can be obtained through diagonalization of the whole family of the effective Hamiltonians.\cite{safkk} 

First, let us discuss the basic tenets of the SES-CC formalism. 
In the exact wave function limit,
the maximum excitation level $m$ included in the cluster operator $T$  is equal to the number of correlated electrons ($N$) while in the approximate
formulations $m\ll N$. Several  typical examples are 
 CCSD
($m=2$),\cite{purvis82_1910} CCSDT ($m=3$),
\cite{ccsdt_noga,ccsdt_noga_err,scuseria_ccsdt}, and 
CCSDTQ ($m=4$),\cite{Kucharski1991,ccsdtq_nevin}  methods. 
Using the second quantization language, the  $T_k$ components of cluster  operator producing $k$-tuply excitations when acting on the reference function  can be expressed as
\begin{equation}
T_k = \frac{1}{(k!)^2} \sum_{i_1,\ldots,i_k; a_1\ldots a_k} t^{i_1\ldots i_k}_{a_1\ldots a_k} E^{a_1\ldots a_k}_{i_1\ldots i_k} \;,
\label{xex}
\end{equation}
where indices $i_1,i_2,\ldots$ ($a_1,a_2,\ldots$) refer to occupied (unoccupied) spin orbitals in the reference function $|\Phi\rangle$.
The excitation operators $E^{a_1\ldots a_k}_{i_1\ldots i_k} $ are defined through strings of standard creation ($a_p^{\dagger}$) and annihilation ($a_p$)
operators
\begin{equation}
E^{a_1\ldots a_k}_{i_1\ldots i_k}  = a_{a_1}^{\dagger}\ldots a_{a_k}^{\dagger} a_{i_k}\ldots a_{i_1} \;,
\label{estring}
\end{equation}
where creation and annihilation operators satisfy the following anti-commutation rules 
\begin{equation}
[a_p,a_q]_+ =
[a_p^{\dagger},a_q^{\dagger}]_+ = 0    \;, \label{comm1}
\end{equation}
\begin{equation}
[a_p,a_q^{\dagger}]_+ = \delta_{pq} \;.\label{comm2}
\end{equation}
The SES-CC approach is based on the particle-hole (p-h) formalism defined with respect to the reference function $|\Phi\rangle$, where 
quasi-operators $b_p$ and $b_p^{\dagger}$ are defined as
\begin{equation}
b_p =
  \begin{cases}
    a_p            &  {\rm if}   \;\; p \in V\\
    a_p^{\dagger}  &  {\rm if}   \;\; p \in O
  \end{cases}
\end{equation}
and 
\begin{equation}
b_p^{\dagger} =
  \begin{cases}
    a_p^{\dagger}            &  {\rm if}   \;\; p \in V\\
    a_p  &  {\rm if}   \;\; p \in O \;,
  \end{cases}
\end{equation}
where  $O$ and $V$ designate sets of occupied and unoccupied spin orbitals. 
Using the p-h formalism we have 
\begin{equation}
 b_p|\Phi\rangle = 0 \;,
 \label{bpvac}
\end{equation}
and
\begin{equation}
E^{a_1\ldots a_k}_{i_1\ldots i_k}  = b_{a_1}^{\dagger}\ldots b_{a_k}^{\dagger} b_{i_k}^{\dagger} \ldots b_{i_1}^{\dagger}  \;.
\label{estring2}
\end{equation}
Additionally, the $b_p/b_q^{\dagger}$ operators satisfy the same anti-commutation relations as
$a_p/a_q^{\dagger}$ operators, i.e.,
\begin{equation}
[b_p,b_q]_+ =
[b_p^{\dagger},b_q^{\dagger}]_+ = 0    \;, \label{comm1ph}
\end{equation}
\begin{equation}
[b_p,b_q^{\dagger}]_+ = \delta_{pq} \;.\label{comm2ph}
\end{equation}
The  p-h formalism significantly simplifies the analysis of the CC  equations. 
It is also easy to notice that  all excitation operators (\ref{estring})
commute, i.e. for 
\begin{eqnarray}
\hspace*{-0.4cm} E^{a_1\ldots a_k}_{i_1\ldots i_k}  &=& 
a_{a_1}^{\dagger}\ldots a_{a_k}^{\dagger} a_{i_k}\ldots a_{i_1}
= b_{a_1}^{\dagger}\ldots b_{a_k}^{\dagger} b_{i_k}^{\dagger} \ldots b_{i_1}^{\dagger}, \label{prr1} \\
\hspace*{-0.4cm} E^{c_1\ldots c_m}_{j_1\ldots j_m}  &=& 
a_{c_1}^{\dagger}\ldots a_{c_m}^{\dagger} a_{j_m}\ldots a_{j_1}
= b_{c_1}^{\dagger}\ldots b_{c_m}^{\dagger} b_{j_m}^{\dagger} \ldots b_{j_1}^{\dagger},
\label{prr2}
\end{eqnarray}
we have 
\begin{equation}
    [E^{a_1\ldots a_k}_{i_1\ldots i_k}, E^{c_1\ldots c_m}_{j_1\ldots j_m}]=0 \;.
    \label{commxy}
\end{equation}

After substituting   Ansatz (\ref{ccwf}) into the Schr\"odinger equation, one gets the energy-dependent form of the
CC equations:
\begin{equation}
(P+Q) He^T|\Phi\rangle= E (P+Q) e^T |\Phi\rangle \;\;,
\label{schreq}
\end{equation}
where $P$ and $Q$ are projection operators onto the reference function ($P=|\Phi\rangle\langle\Phi|$) and onto excited configurations 
(with respect to $|\Phi\rangle$) generated by
the $T$ operator when acting onto the reference function,
\begin{equation}
Q=\sum_{k=1}^{m}\;\sum_{i_1<i_2<\ldots<i_k; a_1<a_2\ldots <a_k}
 |\Phi_{i_1\ldots i_k}^{a_1 \ldots a_k}\rangle\langle \Phi_{i_1\ldots i_k}^{a_1 \ldots a_k}|  \;,
 \label{qoper}
\end{equation}
where 
\begin{equation}
|\Phi_{i_1\ldots i_k}^{a_1 \ldots a_k}\rangle = 
E^{a_1\ldots a_k}_{i_1\ldots i_k} |\Phi\rangle \;.
\label{exsl}
\end{equation}
Diagrammatic analysis \cite{paldus07} leads to an equivalent ({\it at the solution}),  energy-independent form of the CC equations for cluster amplitudes
\begin{equation}
Qe^{-T}He^T|\Phi\rangle = Q(He^T)_C|\Phi\rangle = 0 \;,
\label{conform}
\end{equation}
and energy expression 
\begin{equation}
E=\langle\Phi|e^{-T} H e^T |\Phi\rangle
=\langle\Phi|(H e^T)_C |\Phi\rangle\;,
\label{eneex}
\end{equation}
where $C$ designates a connected part of a given operator expression. In the forthcoming discussion, we refer to 
$e^{-T}He^T$ as a similarity transformed Hamiltonian $\bar{H}$.

The SES-CC formalism hinges upon the notion of excitation sub-algebras of algebra $\mathfrak{g}^{(N)}$ generated by   
$E^{a_l}_{i_l}=b_{a_l}^{\dagger} b_{i_l}$ operators in the particle-hole  representation defined  with respect to the reference $|\Phi\rangle$. As a consequence all generators 
commute, i.e., $[E^{a_l}_{i_l},E^{a_k}_{i_k}]=0$ and algebra $\mathfrak{g}^{(N)}$  (along with all sub-algebras considered here) is commutative.
The SES-CC formalism utilizes an important class of sub-algebras  of commutative
$\mathfrak{g}^{(N)}$ algebra,  which contain all possible excitations
$E^{a_1\ldots a_m}_{i_1\ldots i_m}$ that excite electrons from a subset of active occupied orbitals (denoted as $R$)
to a subset of active virtual orbitals (denoted as $S$).
These sub-algebras will be designated as $\mathfrak{g}^{(N)}(R,S)$.
In the following discussion, we will use $R$ 
and $S$ 
notation for subsets of occupied and virtual active orbitals $\lbrace R_i, \; i=1,\ldots,x \rbrace$ and 
$\lbrace S_i, \; i=1,\ldots,y \rbrace$, respectively (sometimes it is convenient to use alternative notation
$\mathfrak{g}^{(N)}(x_R,y_S)$ where numbers of active orbitals in $R$ and $S$ orbital  sets, $x$ and $y$, respectively,  are explicitly called out). 
Of special interest in building various approximations are sub-algebras that include all $n_v$ virtual orbitals ($y=n_v$) - these sub-algebras will be denoted as 
$\mathfrak{g}^{(N)}(x_R)$.
As discussed in Ref.\cite{safkk} configurations  generated by elements of $\mathfrak{g}^{(N)}(x_R,y_S)$  along with the reference function 
span the complete active space (CAS) referenced to as the CAS($R,S$)
(or eqivalently CAS($\mathfrak{g}^{(N)}(x_R,y_S)$)).

Each sub-algebra  $\mathfrak{h}=\mathfrak{g}^{(N)}(x_R,y_S)$ induces partitioning of the cluster operator $T$  into internal 
($T_{\rm int}(\mathfrak{h}$) or $T_{\rm int}$ for short) part belonging to $\mathfrak{h}$ and external 
($T_{\rm ext}(\mathfrak{h}$) or $T_{\rm ext}$ for short) part not belonging to $\mathfrak{h}$, i.e., 
\begin{equation}
T=T_{\rm int}(\mathfrak{h})+T_{\rm ext}(\mathfrak{h}) \;.
\label{decni}
\end{equation}
In Ref.\cite{safkk}, it was shown that if the  two following  criteria are met:
(1) the $|\Psi(\mathfrak{h})\rangle= e^{T_{\rm int}(\mathfrak{h})}|\Phi\rangle$ is characterized by the same symmetry properties as 
$|\Psi\rangle$ and $|\Phi\rangle$ vectors
(for example,  spin and spatial symmetries),
and (2) the $e^{T_{\rm int}(\mathfrak{h})}|\Phi\rangle$ Ansatz generates FCI  expansion for the sub-system
defined by the CAS corresponding to the $\mathfrak{h}$ sub-algebra,
then $\mathfrak{h}$ is called a sub-system embedding sub-algebra (SES) for cluster operator $T$.
For any SES $\mathfrak{h}$ we proved the equivalence of two representations of the CC equations at the solution:
(i) {\em standard}
\begin{eqnarray}
\langle\Phi| \bar{H}|\Phi\rangle &=& E \;,\label{rr00} \\
Q_{\rm int} \bar{H}\Phi\rangle &=& 0  \;,\label{rr12} \\
Q_{\rm ext} \bar{H}|\Phi\rangle &=& 0 \;, \label{rr34}
\end{eqnarray}
and (ii)  {\em hybrid}
\begin{eqnarray}
(P+Q_{\rm int}) \bar{H}_{\rm ext} e^{T_{\rm int}}|\Phi\rangle &=& E (P+Q_{\rm int})  e^{T_{\rm int}}|\Phi\rangle \;, \label{pp12} \\
Q_{\rm ext} \bar{H} |\Phi\rangle &=& 0 \;, \label{pp34}
\end{eqnarray}
where 
\begin{equation}
\bar{H}_{\rm ext}=e^{-T_{\rm ext}} H e^{T_{\rm ext}} \;,
\label{heffdef}
\end{equation}
and the two projection operators
$Q_{\rm int}(\mathfrak{h})$  and $Q_{\rm ext}(\mathfrak{h})$  ($Q_{\rm int}$ and $Q_{\rm ext}$ for short) are spanned by all excited configurations
generated by acting with $T_{\rm int}(\mathfrak{h})$ and $T_{\rm ext}(\mathfrak{h})$ onto reference function $|\Phi\rangle$,
respectively. The $Q_{\rm int}$ and $Q_{\rm ext}$ projections operators satisfy the condition
\begin{equation}
Q= Q_{\rm int}+ Q_{\rm ext}\;.
\label{qdec}
\end{equation}
The above equivalence shows that the CC energy can be calculated by diagonalizing non-Hermitian effective Hamiltonian $H^{\rm eff}$ defined as 
\begin{equation}
H^{\rm eff}=(P+Q_{\rm int}) \bar{H}_{\rm ext} (P+Q_{\rm int})\;
\label{heffses}
\end{equation}
in the complete active space  corresponding to {\it  any}  SES of  CC formulation defined by cluster operator $T$, i.e.,
\begin{equation}
    H^{\rm eff}(\mathfrak{h})
    e^{T_{\rm int}(\mathfrak{h})}|\Phi\rangle = 
    E e^{T_{\rm int}(\mathfrak{h})}|\Phi\rangle  \;,
    \forall_{{\rm SES}\; \mathfrak{h}} \;.
\label{seseqh}
\end{equation}
Although the idea of effective Hamiltonians has been intensively explored in the past in many areas of physics and chemistry (see Refs.
\cite{bloch1958theorie,des1960extension,lowdin1963studies,schrieffer1966relation,brandow1967linked,soliverez1969effective,schucan1972effective,jorgensen1975effective,mukherjee1975correlation,jezmonk,kutzelnigg1982quantum,
stolarczyk1985coupled,
mukherjee1986linked,durand1983direct,
durand1987effective,jeziorski1989valence,kaldor1991fock,rittby1991multireference,andersson1990second,andersson1992second,
hirao1992multireference,finley1995applications,
glazek1993renormalization,meissner1995effective,
meissner1998fock,nakano1998analytic,
angeli2001n,angeli2001introduction,
canonical1,
mrcclyakh,bravyi2011schrieffer,sahinoglu2020hamiltonian})
the SES-CC formalism enables on to build effective Hamiltonians using
single-reference formulations. Moreover, it is an inherent feature of the single reference  CC Ansatz, which unlike the multi-reference formulations does not assume that  the wave operator act on the multi-dimensional model space and where the corresponding effective Hamiltonian is diagonalized (as an example see the Bloch wave operator formalism). 
We also believe that Eq.(\ref{seseqh})  may be an  interesting contribution from the point of view of recently explored non-Hermitian extensions of quantum mechanics.
\cite{bender1998real,bender2002complex,mostafazadeh2002pseudo,znojil2009three,bishop2013coupled,bishop2020non}

In contrast to the energy-dependent representation of CC equation (\ref{schreq}), Eqs.(\ref{seseqh}) represent true eigenvalue problems corresponding to 
the same eigenvalue $E$ and $e^{T_{\rm int}(\mathfrak{h})}|\Phi\rangle$ as eigenvectors. 
One should also notice that: (1)  the non-CAS related CC wave function components (referred here as external degrees of freedom)
are integrated out and encapsulated in the form of $H^{\rm eff}$, and (2) the internal part of the wave function,  $e^{T_{\rm int}}|\Phi\rangle$ is fully determined 
by diagonalization of $H^{\rm eff}$ in the corresponding CAS.
Separation of external degrees of freedom in the effective Hamiltonians is a desired feature, especially for building its
reduced-dimensionality representation for quantum computing (QC). However, a factor that impedes the use of SES-CC effective Hamiltonians  in quantum computing  is their non-Hermitian character.
It is also worth mentioning that various CC approximations are characterized by various SESs, which is a unique fingerprint of each standard CC approximaiton.
For example, for the restricted Hartree-Fock (RHF) CC formulations the 
$\mathfrak{g}^{(N)}(1_R,y_S)$ and $\mathfrak{g}^{(N)}(2_R,y_S)$
are SESs for CCSD and CCSDTQ approximations (one should also notice that SES for lower-rank CC approximation is also a SES of higher-rank 
CC approximations, i.e. $\mathfrak{g}^{(N)}(1_R,y_S)$ is also a SES for the CCSDTQ approach). 

\begin{figure}
	\includegraphics[angle=0, width=0.51\textwidth]{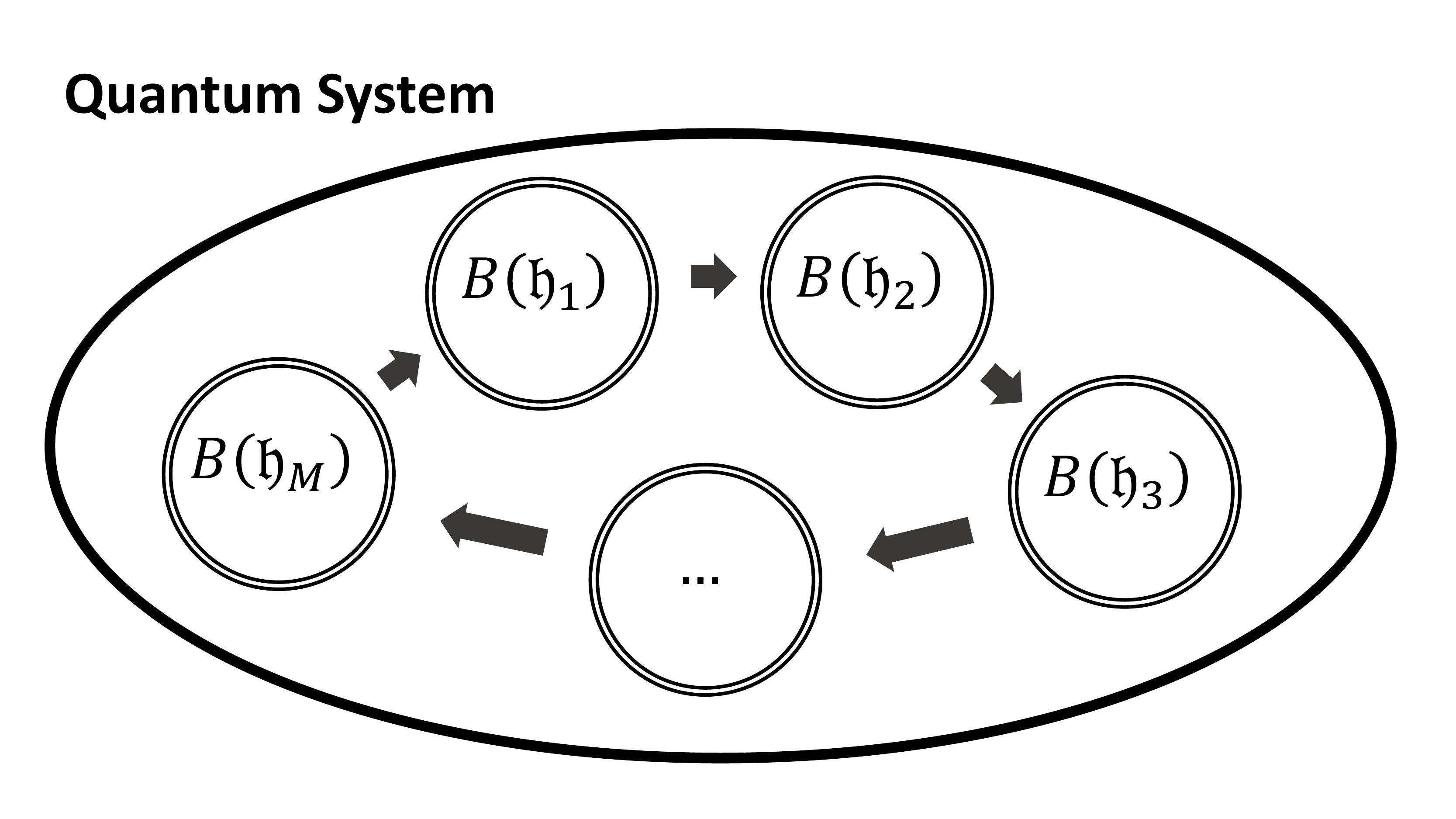}
	\caption{Schematic representation of the CC flow. The entire quantum systems can be probed with various SES-eigenvalue problems (\ref{seseqh}) schematically represented here as $B(\mathfrak{h}_i)$. These computational blocks can be coupled into the flow, where information is passed between various computational blocks $B(\mathfrak{h}_i)$. Subject to the choice of particular classes of SESs defining the flow, the CC flow can probe/traverse a large sub-spaces of entire Hilbert space.}
\label{fig1}
\end{figure}

Properties of SESs-induced eigenvalue problems (\ref{seseqh}) can also be utilized to design new CC approximations based on various amplitude selections processes and re-casting CC equations in a different form, which offer interesting advantages, especially in the way how corresponding equations are solved. This fact can be illustrated on the example of the flow introduced in Ref.\cite{safkk} (see also Fig.\ref{fig1}).
For now (without a loss of generality)  we will focus on special flow involving computational blocks corresponding to $\mathfrak{g}^{(N)}(2_R)$ sub-algebras.
%
While the CCSD equations cannot be  represented as a union of equations corresponding to Eqs.(\ref{seseqh}) for  various CCSD's
SESs  $\mathfrak{g}^{(N)}(1_R,y_S)$ (there are no SES in the CCSD case  that would embrace doubly excited amplitudes 
$t^{ij}_{ab}$ where spinorbitals $i$ and $j$ correspond to distinct orbitals), there exist formalisms  which can probe a significant portion of Hilbert space and are defined by the set of equations that correspond to a union of  non-symmetric eigenvalue problems 
of the type (\ref{seseqh}) for various SESs.
For example, the SCSAF-CCSD(2) approach of Ref.\cite{safkk} uses cluster operator 
$T$ defined as
\begin{equation}
T\simeq T_1+T_2+\sum_{I} T_{{\rm int},3} (\mathfrak{g}^{(N)}(2_{R_I}))
+\sum_{I} T_{{\rm int},4} (\mathfrak{g}^{(N)}(2_{R_I}))
\label{safccsd2}
\end{equation}
where $T_1$ and $T_2$ are singly and doubly excited cluster operators and 
$T_{{\rm int},3} (\mathfrak{g}^{(N)}(2_{R_I}))$ and $T_{{\rm int},4} (\mathfrak{g}^{(N)}(2_{R_I}))$  contain triple and quadruple excitations corresponding to SES $\mathfrak{g}^{(N)}(2_{R_I})$. Summation over $I$ in (\ref{safccsd2})
runs over all possible SESs $\mathfrak{g}^{(N)}(2_{R})$.
It can be shown (see Appendix A)  that in this case, the {\it global} set of CC equations 
\begin{equation}
Q(e^{-T }H e^T)|\Phi\rangle = 0 \;,
\label{bubux}
\end{equation}
where all equations are processed simultaneously in the iterative process of finding the solution, can be re-cast (at the CC solution) in the form of
coupled equations (\ref{seseqh}) of the form 
\begin{equation}
    H^{\rm eff}(\mathfrak{h})
    e^{T_{\rm int}(\mathfrak{h})}|\Phi\rangle = 
    E e^{T_{\rm int}(\mathfrak{h})}|\Phi\rangle  \;,
    \forall_{\mathfrak{h}=\mathfrak{g}^{(N)}(2_{R_I})} \;.
\label{seseqh2}
\end{equation}
%
%
\begin{figure}
	\includegraphics[angle=0, width=0.48\textwidth]{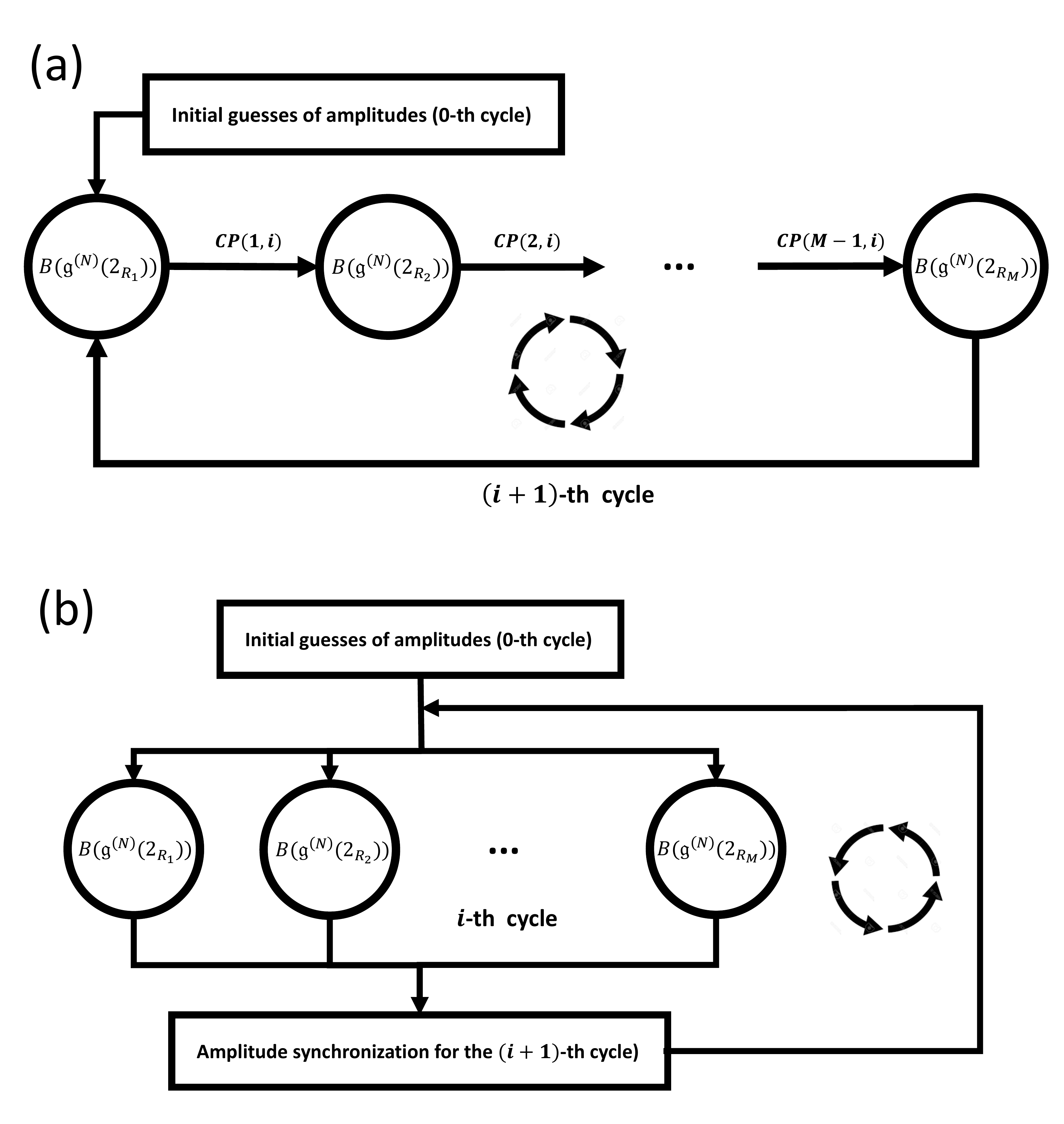}
	\caption{Two types of flow CC formulations for $\mathfrak{g}^{(N)}(2_{R_I})$ sub-algebras (panel (a) represents  serial executions; panel (b) corresponds to the parallel processing of computational blocks (see text for details)).}
\label{fig2}
\end{figure}

As shown in Fig.\ref{fig2}, the solution process of equations (\ref{bubux}) can be organized in the 
form of flow where the algebraic form of  each computational blocks  $B(\mathfrak{g}^{(N)}(2_{R_I}))$  represents 
eigenvalue problem (\ref{seseqh}) for sub-algebra $\mathfrak{g}^{(N)}(2_{R_I})$.
In panel (a) we represent particular flow where particular computational blocks $B(\mathfrak{g}^{(N)}(2_{R_I}))$ are communicating in serial. For this purpose, we first establish an ordering of active spaces defined by $\mathfrak{g}^{(N)}(2_{R_I})$, in the way that reflects their importance  (for example, the first active space contains the most important effects related to the sought for electronic state). Then we define a protocol for passing information between $B(\mathfrak{g}^{(N)}(2_{R_I}))$'s  including  "shared"  cluster amplitudes between various blocks. This problem is caused by the fact that two distinct SESs
$\mathfrak{g}^{(N)}(2_{R_I})$ and 
$\mathfrak{g}^{(N)}(2_{R_J})$ can share a single orbital 
and effectively share all single excitations from this orbital and double excitations exciting $\alpha$ and $\beta$
electrons from the shared orbital.
This redundancy is very small compared to the  total number 
of excitations defining distinct sub-algebras  $\mathfrak{g}^{(N)}(2_{R_I})$ and 
$\mathfrak{g}^{(N)}(2_{R_J})$. For example , there is no overlap with the largest classes of excitations  corresponding to triple and quadruple excitations. 
At the solution,  these redundancies  are irrelevant because the equations for shared amplitudes are the same irrespective of the SESs eigenvalue problem 
(\ref{seseqh2})
they are part of.
To control this effect,
common pool of amplitudes obtained in previous $K$ steps  of $i$-th iteration (denoted as $CP(i,K)$) is 
passed to $K+1$ computational block and external amplitudes (needed to construct $\mathfrak{g}^{(N)}(2_{R_{K+1}})$ effective Hamiltonian)  as well as shared amplitudes 
that correspond to excitations in  $\mathfrak{g}^{(N)}(2_{R_{K+1}})$ enter computational block
$B(\mathfrak{g}^{(N)}(2_{R_{K+1}}))$ as known parameters. 
In this case, the algebraic form of  $B(\mathfrak{g}^{(N)}(2_{R_{K+1}}))$ still takes the form of eigenvalue problem of smaller size 
\begin{widetext}
\begin{eqnarray}
  (P+Q_{\rm int}^{\rm X}) 
  \lbrack
  e^{-T_{\rm int}^{\rm CP}(\mathfrak{h})} 
   H^{\rm eff}(\mathfrak{h})
   e^{T_{\rm int}^{\rm CP}(\mathfrak{h})} 
   \rbrack
    e^{T_{\rm int}^{\rm X}(\mathfrak{h})}|\Phi\rangle &=& 
    E e^{T_{\rm int}^{\rm X}(\mathfrak{h})}|\Phi\rangle  \;,\;\;\;
      \mathfrak{h}=\mathfrak{g}^{(N)}(2_{R_{K+1}}) \;,
 \nonumber \\
   T_{\rm int}(\mathfrak{h}) = T_{\rm int}^{\rm CP}(\mathfrak{h})
   +T_{\rm int}^{\rm X}(\mathfrak{h}) \;,\;\;\;&&
   \mathfrak{h}=\mathfrak{g}^{(N)}(2_{R_{K+1}})\;,
 \label{xcpamp}  
\end{eqnarray}
\end{widetext}
where $T_{\rm int}^{\rm CP}(\mathfrak{h})$ is a part of 
$T_{\rm int}(\mathfrak{h})$  determined by shared amplitudes from the common pool of amplitudes and $T_{\rm int}^{\rm X}(\mathfrak{h})$ is a part of $T_{\rm int}(\mathfrak{h})$ that is determined in the $K+1$ computational blocks.
After Eq.(\ref{xcpamp}) is solved , the $CP(i,K)$ is update for the $T_{\rm int}^{\rm X}(\mathfrak{h})$ amplitudes defining in this way $CP(i,K+1)$ and process is continued with the $K+2$ block. After $M$ steps, the $CP(i,M)$ is used as a
starting common pool of amplitudes for the $i+1$ iteration.
In Eq.(\ref{xcpamp}), the $Q_{\rm int}^{\rm X}$ operator is the projection operator onto excited configurations generated by $T_{\rm int}^{\rm X}(\mathfrak{h})$ when acting onto reference function $|\Phi\rangle$.
In panel (b) we see an alternative "parallel" flow where 
computational blocks are independent and corresponds to the 
original eigenproblems (\ref{seseqh}) for SESs $\mathfrak{g}^{(N)}(2_{R_{I}})$. This step is followed by a sync-up of all shared amplitudes by various SESs. 
More details on the CC flows can be found in Appendices A,B,C, and D, where we discuss the equivalence of global representation and coupled computational blocks involved in the flow, 
the time domain extension, approximate solvers for computational blocks, and general algorithmic structure of the practical CC flow realization. 

The CC flow equations (\ref{seseqh}) or (\ref{seseqh2}) can also be viewed as a {\it configurational (or more aptly - sub-space) version of the Aufbau principle}, which is a consequence of the fact that each problem corresponding to some SES $\mathfrak{h}$
provides a rigorous mechanism for extending the sub-space probed in a flow. In other words, the spaces probed in each SES problem (\ref{seseqh}) are additive.  This fact is a unique feature, which 
should be referred to as the sub-system "memory" of the CC wave function. Additionally, CC flow assures {\em size-consistency} of the calculated ground-state energies.
Similar flows cannot be easily constructed using configuration interaction type methods or standard many-body perturbation techniques. We believe that this is yet another argument in favor of non-perturbative analysis of CC equations. 

In general, flow-based CC formulations are very flexible and allows one to use sub-algebras $\mathfrak{g}^{(N)}(x_R,y_S)$  defined by various $x$ and $y$ parameters. This property may be used to introduce selective groups of higher excitations and tune the cost of flow equations to available computing resources. 
Although the numerical implementation of flow-based formalism may be numerically less efficient than the implementation based on the global representation, its advantage lies in the fact that computational blocks contributing to flow are physically interpretable in terms of Schr\"odinger-type equations for sub-systems described by relevant SESs. 
This fact has profound consequences and allows one to construct more justified (or equivalently less "postulated") and better-controlled approximations based on the flow equations. 
An interesting illustration of this fact will be CC flow  equations for localized orbitals (see Section \ref{s2c}).

\begin{figure}
	\includegraphics[angle=0, width=0.48\textwidth]{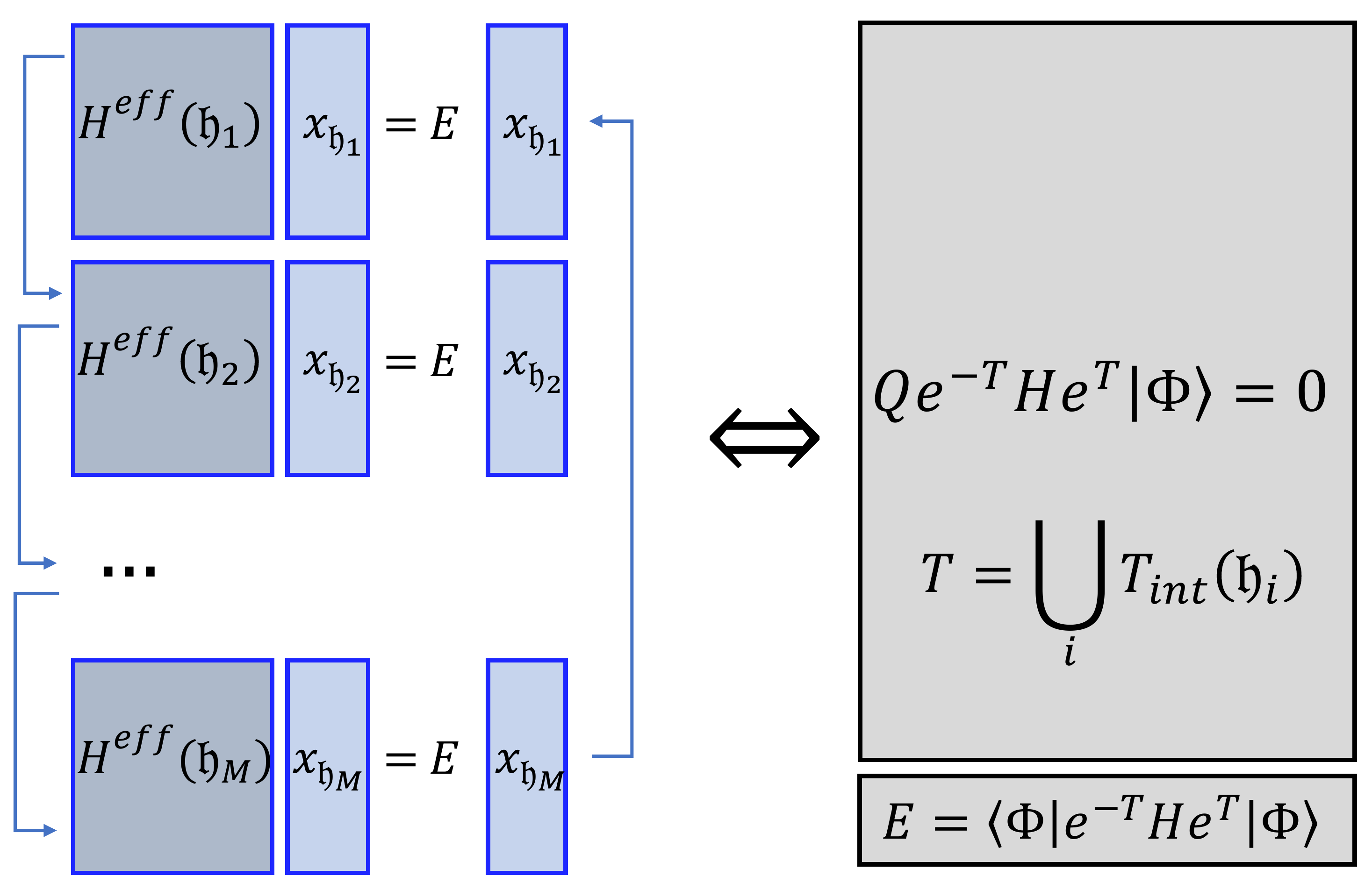}
	\caption{Schematic representation of the equivalence between  the "global" representation of the CC equations and  coupled  SES eigenvalue problems for properly defined cluster operators  (see text for details).}
\label{fig3}
\end{figure}

Summarizing, the flow equations as shown in Fig.\ref{fig3} can be represented in the form of the "global"   connected CC equations with the cluster operator $T$ defined as a union of unique internal excitations of all  sub-systems 
included in the flow. The inverse statement is also true:  for specific choices of the excitation domain included in the cluster operator, the CC equations can be represented in the form of coupled eigenvalue problems corresponding to various 
sub-systems embedding algebras. This statement is a foundation for the reduced-scaling formulations discussed in this paper. 



%

\subsection{Time-dependent CC flows} 
The extension of the CC methods to the time domain keeps attracting much attention in various fields of chemistry and physics. 
Several developments  in this area, including Arponen's seminal papers on this subject \cite{arponen83_311} (see also Refs.\cite{arponen1987extended1,arponen1987extended,
arponen1991independent,arponen1993independent,arponen1993independentx,kvaal2012ab,pedersen2019symplectic,kvaal2020guaranteed,sato2018communication}) 
paved the way for mature applications of these formulations to describe time-dependent physical and chemical processes. 

In Ref.\cite{downfolding2020t}, we demonstrated that the  SES-based downfolding techniques could also be extended to the  time-dependent Schr\"odinger equation when all orbitals and the reference functions $|\Phi\rangle$ are assumed to be time-independent. As in the stationary case,  we will assume a general partitioning of the 
time-dependent cluster operator $T(t)$ into its
internal ($T_{\rm int}(\mathfrak{h},t)$) and external ($T_{\rm ext}(\mathfrak{h},t)$) parts, i.e,
\begin{equation}
|\Psi(t)\rangle 
= e^{T_{\rm ext}(\mathfrak{h},t)} e^{T_{\rm int}(\mathfrak{h}, t)} |\Phi\rangle \;,
\forall \mathfrak{h} \in SES  \;. \label{step2} 
\end{equation}
For generality, we also include phase factor $T_0(\mathfrak{h},t)$ in the definition of the  $T_{\rm int}(\mathfrak{h},t)$ operator.
After substituting (\ref{step2}) into time-dependent  Schr\"odinger equation and utilizing properties of SES algebras, we demonstrated that the ket-dynamics of the sub-system wave function 
$e^{T_{\rm int}(\mathfrak{h},t)}|\Phi\rangle$
corresponding to arbitrary SES $\mathfrak{h}$
\begin{equation}
i\hbar \frac{\partial }{\partial t} e^{T_{\rm int}(\mathfrak{h},t)} |\Phi\rangle = H^{\rm eff}(\mathfrak{h},t)  e^{T_{\rm int}(\mathfrak{h},t)} |\Phi\rangle \;,
\label{cool2}
\end{equation}
where 
\begin{equation}
H^{\rm eff}(\mathfrak{h},t) = (P+Q_{\rm int}(\mathfrak{h})) \bar{H}_{\rm ext}(\mathfrak{h},t) (P+Q_{\rm int}(\mathfrak{h})) \;
\label{lemma2}
\end{equation}
and 
\begin{equation}
\bar{H}_{\rm ext}(\mathfrak{h},t) = e^{-T_{\rm ext}(\mathfrak{h},t)} H e^{T_{\rm ext}(\mathfrak{h},t)} \;.
\label{hbart}
\end{equation}
In analogy to the stationary cases, various sub-systems  computational blocks can be integrated into a flow enabling sampling of large sub-spaces of Hilbert space 
through a number of coupled reduced-dimensionality  problems. For example, the time-dependent variant of the 
the SCSAF-CCSD(2) approach uses the time-dependent  cluster operator 
$T(t)$ in the form 
\begin{eqnarray}
T(t)&\simeq& T_1(t)+T_2(t)+\sum_{I} T_{{\rm int},3} (\mathfrak{g}^{(N)}(2_{R_I}),t)
\nonumber \\
&&+\sum_{I} T_{{\rm int},4} (\mathfrak{g}^{(N)}(2_{R_I}),t))
\label{safccsd2xx}
\end{eqnarray}
which can be equivalently represented as 
coupled time-evolution problems for  sub-systems
\begin{equation}
i\hbar \frac{\partial }{\partial t} e^{T_{\rm int}(\mathfrak{h},t)} |\Phi\rangle = H^{\rm eff}(\mathfrak{h},t)  e^{T_{\rm int}(\mathfrak{h},t)} |\Phi\rangle \;,
\forall_{\mathfrak{h}=\mathfrak{g}^{(N)}(2_{R_I})} \;,
\label{seseqht}
\end{equation}
(for details see Appendix B). 
These equations can be solved using similar flows as shown in Fig.\ref{fig2} with the difference that now the iterative cycles for converging amplitudes/energy correspond to elementary time steps with increment corresponding to $\Delta t$. 
As in the stationary case, the time-dependent CC flow equations represent an extension  of the "sub-space Aufbau" principle mentioned earlier, where each SES-problem (\ref{seseqht}) extends the space probed in time-dependent CC formalism. 
In the view of deep analogies between stationary CC flow equations based on the localized orbitals and local CC formulations developed in the last few decades in quantum chemistry (see the next Subsection), 
the flow described by Eq.(\ref{seseqht}) 
can be considered as a reduced-scaling variant of the time-dependent CC formulations.

\subsection{CC flows for localized orbital basis and localized sub-systems} 
\label{s2c}
The SES-CC flow  formalism also 
systematizes and further extends the notion of "sub-system" composed of orbital pairs. This problem has intensively been studied in early non-orthogonal pseudo-natural orbitals based formulations of CI \cite{meyer1973pno},  coupled electron pair approximation (CEPA) \cite{meyer1975pno}, and their extensions to local CEPA/CC methods based on the local pair natural orbitals (LPNO) and their 
domain-based LPNO variant (DLPNO).\cite{neese2009efficient,neese2009accurate}
To analyze SES-CC approximations, let us, in the analogy to the DLPNO-CC formulations, assume that the set of selected orbital pairs (for simplicity, we will focus on the closed-shell formulations)  
$\cal{P}$=$\lbrace (i,j) \rbrace$
that significantly contribute to correlation energy is known.  These pairs are also employed to define PNO spaces and corresponding  CCSD cluster amplitudes. 
In the standard pair-driven DLPNO-CCSD approximation 
orbital pairs $(i,j)$ (including pairs where $i=j$, i.e., $(i,i)$) along with $(i,j)$-specific natural virtual orbitals are used to select PNO space and relevant single and double excitations. The $(i,j)$-specific density matrix defined as MP2-type density matrix ${\bf D}^{ij}$ (see Ref.\cite{riplinger2013efficient})
 is used to determine virtual PNOs in the way that only natural orbitals characterized by occupation numbers greater than the user-defined threshold are retained. 
It leads to a significant reduction in the size of  pair-specific PNOs  spaces and consequently to a significant reduction of the number of pair-specific singly ($t^i_{a_{ii}}$) and doubly ($t^{ij}_{a_{ij} b_{ij}}$) excited amplitudes, where virtual indices $a_{ii}$, $a_{ij}$, and $b_{ij}$ are defined by reduced-size PNOs.
It should be noted that each pair $(i,j)$ introduces its own set of PNO virtual orbitals, which may not be orthogonal to PNOs corresponding to distinct pairs. 

The CC $\mathfrak{g}^{(N)}(2_R)$ flow employing localized occupied orbitals in a natural way introduces several elements underlying 
standard DLPNO-CCSD  design. Our analysis is based on the observation that
there is a a natural correspondence between orbital
pairs from $\cal{P}$ (along with  all virtual orbitals) 
with $\mathfrak{g}^{(N)}(2_R)$ sub-algebras, where
\begin{equation}
    R\equiv (i,j) \;,\;\; (i,j)\in \cal{P}\;.
    \label{rij}
\end{equation}
For short  we will refer to  these SESs as $\mathfrak{g}^{(N)}(2_{ij})$ ($(i,j)\in \cal{P} $).
Additionally,  each Schr\"odinger-type equation in  (\ref{seseqh}) or (\ref{seseqh2})
\begin{equation}
    H^{\rm eff}(\mathfrak{h})
    e^{T_{\rm int}(\mathfrak{h})}|\Phi\rangle = 
    E e^{T_{\rm int}(\mathfrak{h})}|\Phi\rangle \;,
    \mathfrak{h}= \mathfrak{g}^{(N)}(2_{ij})\;\; (i,j)\in \cal{P}\;.
\label{seseqhloc}
\end{equation}
naturally defines corresponding   one-body density matrix
$\rho(\mathfrak{g}^{(N)}(2_{ij}))$ 
and its  PNOs (or DLPNOs) without any additional assumptions regarding the form and the origin of the pair density matrix. 
One can readily notice that the ${\bf D}^{ij}$ density matrices used in original DLPNO-CC papers is its low-order  approximation. 
This feature of SES equations (\ref{seseqhloc}) can be viewed as a CC-derived systematization of  the sub-system  (or  pair) concepts  discussed in original works of Sinano\u{g}lu 
\cite{sinanouglu1962many,sinanoglu1964many} and Meyer \cite{meyer1971ionization,meyer1973pno,meyer1975pno}.

It is worth mentioning that in contrast to the DLPNO-CCSD formalism, the $\mathfrak{g}^{(N)}(2_{ij})$ include specific classes of triple and quadruple excitations. This feature may be a possible way to define the balanced inclusion of higher-rank excitations in the local CC formulations.
For this purpose one can also envision CC flows based on $\mathfrak{g}^{(N)}(x_R)$ sub-algebras with $x > 2$ that employ localized orbitals. 

Since each SES $(\mathfrak{g}^{(N)}(2_{ij}))$  computational block contributing to flow (\ref{seseqhloc}) defines its own set of PNOs, in analogy to DLPNO-CC approaches,  it can be re-expressed  through its own set of PNOs and set of pre-selected internal amplitudes   (utilizing predefined thresholds). 
However, introducing threshold in the flow CC equations can be performed less abruptly than in the existing DLPNO-CC methods. In fact, the amplitude selection process is equivalent of selecting sub-algebra $\mathfrak{g}^{(N)}(2_{ij},y_S)$ of 
$\mathfrak{g}^{(N)}(2_{ij})$ in the way that  $S$ is the set of PNOs corresponding to occupation number greater than the predefined threshold and $y$ is the total number of virtual PNOs  selected this way.
This selection induces a natural partitioning of the $T_{\rm int}(\mathfrak{g}^{(N)}(2_{ij}))$ into  a
part belonging to $\mathfrak{g}^{(N)}(2_{ij},y_S)$
($T_{\rm int}(\mathfrak{g}^{(N)}(2_{ij},y_S))$ and remaining "neglegible" part of excitations
($\Delta T_{\rm int}(\mathfrak{g}^{(N)}(2_{ij}))$)
\begin{equation}
    T_{\rm int}(\mathfrak{g}^{(N)}(2_{ij})) = 
    T_{\rm int}(\mathfrak{g}^{(N)}(2_{ij},y_S))
    +\Delta T_{\rm int}(\mathfrak{g}^{(N)}(2_{ij})) \;\;.
    \label{intdec1}
\end{equation}
The controlled version of the selection step is achieved by noticing that 
the effect of 
 the "neglegible" amplitudes $\Delta T_{\rm int}(\mathfrak{g}^{(N)}(2_{ij}))$ can still be absorbed (using another downfolding step within the  $ \mathfrak{g}^{(N)}(2_{ij})$ CAS space) in the form of additional similarity transformation:
\begin{widetext}
\begin{equation}
  (P+Q_{\rm int}(\mathfrak{f}_{ij}))
  \lbrack
 e^{-\Delta T_{\rm int}(\mathfrak{h}_{ij})}
 H^{\rm eff}(\mathfrak{h}_{ij})
e^{\Delta T_{\rm int}(\mathfrak{h}_{ij})} 
  \rbrack
  e^{T_{\rm int}(\mathfrak{f}_{ij})}|\Phi\rangle = 
   E 
   e^{T_{\rm int}(\mathfrak{f}_{ij})}|\Phi\rangle
    \;,\; \mathfrak{f}_{ij}=\mathfrak{g}^{(N)}(2_{ij},y_S) \;,\;
    \mathfrak{h}_{ij}=\mathfrak{g}^{(N)}(2_{ij})
\label{seseqh3}
\end{equation}
\end{widetext}     
where $Q_{\rm int}(\mathfrak{g}^{(N)}(2_{ij},y_S))$ is a projection operator onto excited configurations generated by the 
$T_{\rm int}(\mathfrak{g}^{(N)}(2_{ij},y_S))$ when acting onto reference function $|\Phi\rangle$.
%

The CC flow equations can also be naturally linked to approximate CC schemes used in studies of spin systems.
In Refs.\cite{bishop1991coupled,bishop2011coupled}  Bishop {\it et al.} considered a hierarchy of approximations
 where excitation manifolds are defined using the so-called sub-systems defined by contiguous lattice sites, each of which is nearest neighbors to at least one other in the sub-system. 
These sub-systems can be naturally identified with the active spaces, making CC flows equations similar to the SUB$n$-$m$ and LSUB$m$ schemes discussed in Refs.\cite{bishop1991coupled,bishop2011coupled}
As shown in the previous paragraphs, the application of CC flows to the spin systems offers an interesting way of defining reduced-scaling methods for quantum lattice models. 
This can be achieved by selecting the essential class of excitations (for a given sub-system) using density matrices corresponding to sub-system's  downfolded Hamiltonians. 
This approach can address problems associated with high numerical overheads of highly accurate SUB$n$-$m$ and LSUB$m$ approaches. 
The related developments will be discussed in forthcoming papers. 

Summarizing,  several basic threads of DLPNO-CCSD equations are consequences of CC flow equations defined by $\mathfrak{g}^{(N)}(2_{ij})$ sub-algebras. The local character of correlation effects is a net effect of the local character of the basis set used, asymptotic properties of one- and two-electron interactions, and fundamental properties of CC formalism associated with the CC  sub-system memory. This feature allows one to construct, in a rigorous way,  Schr\"odinger-type  equations for sub-systems (in this case, a pair of orbitals) defining the flow. 
Additionally, a rigorous definition of sub-system and associated wave function lead to a natural definition of the sub-system density matrix and natural orbitals. 
We believe that the CC flow equations based on the SES formalism are an interesting tool for constructing various approximations for correlated systems. This formulation can be universally used in stationary formulations of canonical and local CC formulations and be extended to the time-dependent CC equations.  
The general CC flow formalism is not  limited to types of interactions that are  considered molecular systems  and can also be extended to other types of many-body interactions encountered in chemical and physical applications.


\section{Sub-system flows based on the double unitary CC representations of wave function}
We find properties of SR-CC flows very appealing from the point of view of quantum computing. Instead of considering expensive  "global" space approaches (as done in the majority of existing QC formalisms) that require too many parameters to be optimized at the same time, one could partition the problem into smaller computational sub-problems that can be tuned to available systems of qubits. 
For this reason, we would like to adapt the SR-CC ideas from  previous sections to double unitary CC Ansatz (DUCC; see Ref.\cite{bauman2019downfolding}).
While the DUCC formalism  mirrors some properties of the  SES-CC formalism and additionally assures the Hermitian character of the 
effective Hamiltonians in CAS($\mathfrak{h}$), due to the non-commutative nature of the anti-Hermitian cluster operators 
employed by this formalism,
coupling various DUCC problems into a flow requires several approximations, described in the following subsection. 

The DUCC formalism discussed in Refs.\cite{bauman2019downfolding,downfolding2020t} uses a composite 
unitary CC Ansatz  to represent the exact wave function $|\Psi\rangle$, i.e.,
\begin{equation}
        |\Psi\rangle=e^{\sigma_{\rm ext}(\mathfrak{h})} e^{\sigma_{\rm int}(\mathfrak{h})}|\Phi\rangle \;,
\label{ducc1}
\end{equation}
where $\sigma_{\rm ext}(\mathfrak{h})$ and $\sigma_{\rm int}(\mathfrak{h})$ are general-type anti-Hermitian operators
\begin{eqnarray}
\sigma_{\rm int}^{\dagger}(\mathfrak{h}) &=&  -\sigma_{\rm int}(\mathfrak{h}) \;, \label{sintah} \\
\sigma_{\rm ext}^{\dagger}(\mathfrak{h}) &=&  -\sigma_{\rm ext}(\mathfrak{h}) \;. \label{sintah2}
\end{eqnarray} 
All cluster amplitudes defining $\sigma_{\rm int}$ cluster operator carry active indices only (or indices of active orbitals defining given $\mathfrak{h}$). The external part $\sigma_{\rm ext}(\mathfrak{h})$ is defined by amplitudes carrying at least one inactive orbital index. In contrast to the SR-CC approach,  internal/external parts of anti-Hermitian cluster operators are not defined in terms of excitations belonging explicitly to a given sub-algebra but rather by indices defining active/inactive orbitals specific to a given $\mathfrak{h}$.
Therefore $\mathfrak{h}$ will be used here in the context of CAS's generator. 

When the external cluster amplitudes are known (or can be effectively approximated), in analogy to single-reference SES-CC formalism, the energy (or its approximation) can be calculated by diagonalizing Hermitian effective/downfolded Hamiltonian in the active space using various quantum or classical diagonalizers. 
An important step towards developing practical computational schemes is to simplify the infinite expansions defining both cluster amplitudes and non-terminating commutator expansions defining downfolded Hamiltonians. 
A legitimate approximation of $\sigma_{\rm ext}(\mathfrak{h})$ and $\sigma_{\rm int}(\mathfrak{h})$
in Eq.(\ref{ducc1})  for well-defined active spaces is to retain  lowest-order terms only, i.e.,
\begin{eqnarray}
\sigma_{\rm int}(\mathfrak{h}) &\simeq& T_{\rm int}(\mathfrak{h}) - T_{\rm int}(\mathfrak{h})^{\dagger} \;, \label{sint} \\
\sigma_{\rm ext}(\mathfrak{h}) &\simeq& T_{\rm ext}(\mathfrak{h}) - T_{\rm ext}(\mathfrak{h})^{\dagger} \;, \label{sext}
\end{eqnarray}
which has been discussed in Ref.\cite{bauman2019downfolding}. 
In particular, $T_{\rm ext}(\mathfrak{h})$ can be approximated by SR-CCSD amplitudes that carry at least one external spinorbital index. Other possible sources for obtaining external cluster amplitudes are higher-rank SR-CC methods and approximate unitary CC formulations such as UCC(n) methods \cite{unitary1,unitary2}.

Using DUCC representation (\ref{ducc1}) it can be shown that in analogy to the SR-CC case, the energy of the entire system (once the exact form of $\sigma_{\rm ext}(\mathfrak{h})$ operator is known) can be calculated through the diagonalization of the effective/downfolded Hamiltonian in SES-generated active space, i.e., 
\begin{equation}
        H^{\rm eff}(\mathfrak{h}) e^{\sigma_{\rm int}(\mathfrak{h})} |\Phi\rangle = E e^{\sigma_{\rm int}(\mathfrak{h})}|\Phi\rangle,
\label{duccstep2}
\end{equation}
where
\begin{equation}
        H^{\rm eff}(\mathfrak{h}) = (P+Q_{\rm int}(\mathfrak{h})) \bar{H}_{\rm ext}(\mathfrak{h}) (P+Q_{\rm int}(\mathfrak{h}))
\label{equivducc}
\end{equation}
and 
\begin{equation}
        \bar{H}_{\rm ext}(\mathfrak{h}) =e^{-\sigma_{\rm ext}(\mathfrak{h})}H e^{\sigma_{\rm ext}(\mathfrak{h})}.
\label{duccexth}
\end{equation}
%
%
\begin{figure}
	\includegraphics[angle=0, width=0.48\textwidth]{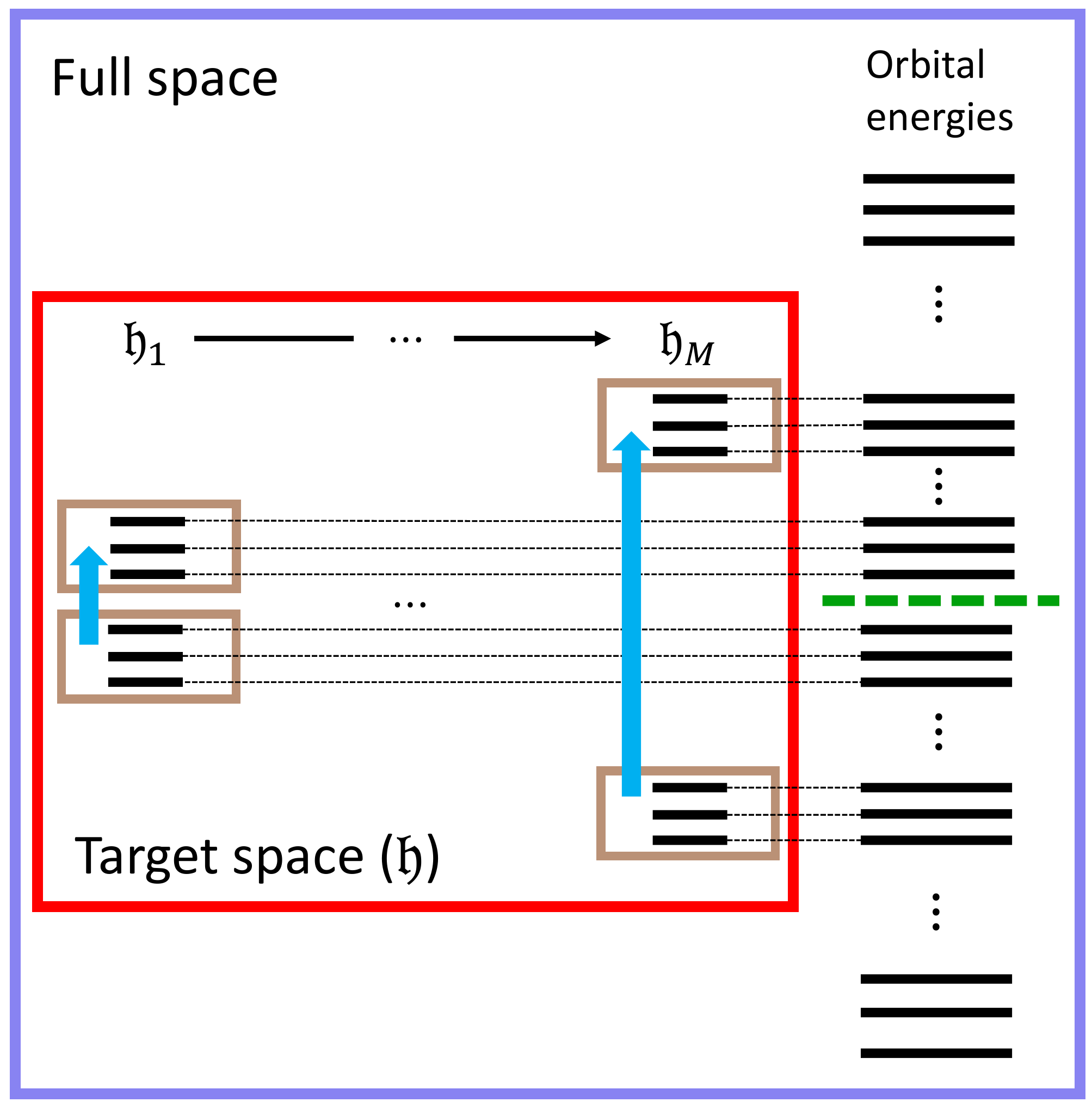}
	\caption{Schematic representation of the DUCC flow. 
	It is assumed that the most important classes of excitations required to describe a state of interest are captured by the target active space (or virtual space)  too large for direct QC simulations and that  the target active space (corresponding to sub-algebra $\mathfrak{h}$) can be "approximated" by excitations included 
	in smaller yet  computationally feasible active spaces corresponding to 
	sub-algebras $\mathfrak{h}_1$, ..., $\mathfrak{h}_M$ (see text).  The DUCC flow combines 
	computational blocks that correspond to  variational problems associated with
	each sub-algebra $\mathfrak{h}_i \; (i=1,\ldots,M)$.
	The green dashed line represents Fermi level.}
\label{fig4}
\end{figure}
Typical  approximations for downfolded Hamiltonian utilize: (1) various sources for evaluation of the
$T_{\rm ext}(\mathfrak{h})$ operator in (\ref{sext}), (2) 
various length of commutator expansion defining the 
$\bar{H}_{\rm ext}(\mathfrak{h})$ operator, (3) various excitation-ranks in the many-body expansion of the 
$\bar{H}_{\rm ext}(\mathfrak{h})$ operator, and (4) various molecular basis choices.

Recently, applications of QPE and VQE quantum  algorithms to evaluate eigenvalues of  downfolded Hamiltonians $H^{\rm eff}(\mathfrak{h})$ became a subject of intensive studies. 
In the case of the VQE method, the  energy functional 
\begin{equation}
\min_{{\bm \theta}(\mathfrak{h})}   
\langle\Psi({\bm \theta}(\mathfrak{h}))|H^{\rm eff}(\mathfrak{h})|
\Psi({\bm \theta}(\mathfrak{h}))\rangle
\label{vqeducc}
\end{equation}
is optimized with respect to variational parameters 
${\bm \theta}(\mathfrak{h})$
where $|\Psi({\bm \theta}(\mathfrak{h}))\rangle$ approximates 
$e^{\sigma_{\rm int}(\mathfrak{h})}|\Phi\rangle$
\begin{equation}
   |\Psi({\bm \theta}(\mathfrak{h}))\rangle \simeq  e^{\sigma_{\rm int}(\mathfrak{h})}|\Phi\rangle
   \label{vqeapp}
\end{equation}
at the level of  quantum circuit.
This approach turned out to be very efficient, especially when "correlated" natural orbitals are employed. The advantage of using the VQE approach is the possibility of extracting the information 
about cluster amplitudes defining 
$\sigma_{\rm int}(\mathfrak{h})$ from the  optimized parameters ${\bm \theta}(\mathfrak{h})$.
This feature  plays a vital role in designing DUCC sub-system flows and assures mechanism of quantum information passing between various computational blocks.
In the following analysis we will assume that the variational parameters ${\bm \theta}(\mathfrak{h})$ correspond to   cluster amplitudes  in  $\sigma_{\rm int}(\mathfrak{h})$ expansion. 

%
%
%
%
%

\subsection{DUCC flow equations: applications in quantum computing}

The DUCC flow idea is very interesting from the point of view of its applications in quantum computing, where a quantum computer can process computational blocks (either corresponding to energy functional minimization or diagonalization of the downfolded Hamiltonians). 
In this section, we will extend the idea of the SR-CC sub-system flow to the DUCC formalism, and we will highlight the similarities and differences between these two approaches. 
The main differences between SR-CC and DUCC should be attributed to the non-commutative nature of many-body components defining anti-Hermitian DUCC cluster operators. This fact, in the case of the DUCC approach,  significantly impedes the analysis of the equations and partitioning them into separate computational blocks that can be integrated into a sub-system flow equations. However, this can be achieved with a sequence of approximations we describe below.  

We will start our analysis from assuming that we would like to perform DUCC effective simulations for SES $\mathfrak{h}$ problem (\ref{duccstep2})
which is, for whatever reason, too complex or too big for quantum  processing. We will assume that external amplitudes $\sigma_{\rm ext}(\mathfrak{h})$
can be effectively evaluated using perturbative formulations. For simplicity we will introduce a new DUCC Hermitian Hamiltonian $A(\mathfrak{h})$ which is defined 
as $H^{\rm eff}(\mathfrak{h})$ or its approximation in the 
$(P+Q(\mathfrak{h}))$ space (in the simplest case it can be just 
the $(P+Q(\mathfrak{h}))H(P+Q(\mathfrak{h}))$ operator). 
We will denote $A(\mathfrak{h})$ simply by $A$.
We will also   assume the situation where   excitations  from  $\mathfrak{h}$ that are relevant 
to state of interest can be captured by excitation sub-algebras: $\mathfrak{h}_1$, $\mathfrak{h}_2$, $\ldots$, 
$\mathfrak{h}_M$ (see Fig.\ref{fig4}), where, in analogy to the SR-CC case, we admit the possibility of "sharing" excitations/de-excitations between these sub-algebras. We also assume that the number of excitations belonging to  each 
$\mathfrak{h}_i$ $(i=1,\ldots,M)$ is significantly smaller  than the number of excitations in $\mathfrak{h}$ and therefore 
numerically tractable in quantum simulations. Below we will discuss the challenges and approximations that are needed to obtain well defined
DUCC flow equations. 

The  $A(\mathfrak{h})$ Hamiltonian and the
$(P+Q(\mathfrak{h}))$ space can be treated as a starting point for the secondary DUCC  decompositions
generated by sub-system algebras
$\mathfrak{h}_i$ $(i=1,\ldots,M)$
defined above,
i.e.,
\begin{equation}
    A^{\rm eff}(\mathfrak{h}_i)
    e^{\sigma_{\rm int}(\mathfrak{h}_i)}|\Phi\rangle = 
    E e^{\sigma_{\rm int}(\mathfrak{h}_i)}|\Phi\rangle \;
   \;(i=1\ldots,M)\;.
\label{seseqh22}
\end{equation}
or in the VQE-type variational representation as 
\begin{equation}
\min_{{\bm \theta}(\mathfrak{h}_i)}   
\langle\Psi({\bm \theta}(\mathfrak{h}_i))|A^{\rm eff}(\mathfrak{h}_i)|
\Psi({\bm \theta}(\mathfrak{h}_i))\rangle \;\;
(i=1\ldots,M)\;.
\label{vqeduccx1}
\end{equation}
Each $A^{\rm eff}(\mathfrak{h}_i)$ is defined as
\begin{equation}
        A^{\rm eff}(\mathfrak{h}_i) = (P+Q_{\rm int}(\mathfrak{h}_i)) \bar{A}_{\rm ext}(\mathfrak{h}_i) (P+Q_{\rm int}(\mathfrak{h}_i))
\label{equivducc66}
\end{equation}
and 
\begin{equation}
        \bar{A}_{\rm ext}(\mathfrak{h}_i) =e^{-\sigma_{\rm ext}(\mathfrak{h}_i)}A  e^{\sigma_{\rm ext}(\mathfrak{h}_i)}.
\label{duccexth89}
\end{equation}
where we defined external $\sigma_{\rm ext}(\mathfrak{h}_i)$ operator
with respect to $\mathfrak{h}$ or
$(P+Q_{\rm int}(\mathfrak{h}))$ space (i.e.cluster amplitudes defining $\sigma_{\rm ext}(\mathfrak{h}_i)$ must carry at last one index belonging do active spin orbitals defining $\mathfrak{h}$ and not belonging to 
set of active spin orbtitals defining $\mathfrak{h}_i$).
In other words, sub-algebras $\mathfrak{h}_i$  generate active sub-spaces in larger 
active space $\mathfrak{h}$, i.e.,
$(P+Q_{\rm int}(\mathfrak{h}_i)) \subset (P+Q(\mathfrak{h}))$. 
However, connecting  DUCC computational blocks (\ref{seseqh22}) or (\ref{vqeduccx1}) directly into a flow is a rather challenging task. In contrast to the SR-CC sub-system flows where cluster amplitudes are universal for all sub-algebras induced problems (i.e., given amplitude carries the same value across all computational blocks), the same is no longer valid for DUCC flows.
Again, this is a consequence of the non-commutativity of the anti-Hermitian operators defining DUCC representation of the wave function. For example, the internal amplitude for
some $\mathfrak{h}_i$ problem may assume a different value as the same amplitude being an internal 
amplitude for a different problem corresponding to sub-algebra 
$\mathfrak{h}_j$ $(i\ne j)$, which means that DUCC amplitudes explicitly depend on the sub-algebra index $\mathfrak{h}_i$
(as opposed to the SR-CC flow formalism, where values of particular amplitudes were independent of the sub-algebra index).
Similar effects could be observed in ADAPT-VQE formulations,\cite{grimsley2019adaptive} where amplitudes from the excitation poll may appear multiple times carrying various values in the wavefunction expansion. 
An additional problem is related to the fact that while for the SR-CC flows, effective Hamiltonians corresponding for various SESs can be constructed exactly, for the DUCC case $A^{\rm eff}(\mathfrak{h}_i)$ can be constructed only in an approximate way, and therefore their  ground-state  eigenvalues may not be exactly  equal. 

To address these issues and define practical DUCC flow we will discuss the algorithm that combines secondary downfolding steps with Trotterization of the unitary CC operators. 
Let us  assume that the $\sigma_{\rm int}(\mathfrak{h})$ operator can be approximated by $\sigma_{\rm int}(\mathfrak{h}_i)
 (i=1,\ldots,M)$, i.e.,
\begin{equation}
    \sigma_{\rm int}(\mathfrak{h}) \simeq
    \sum_{i=1}^M \sigma_{\rm int}(\mathfrak{h}_i) +
    X(\mathfrak{h},\mathfrak{h}_1,\ldots,\mathfrak{h}_M)
    \label{trott1}
\end{equation}
where the  $X(\mathfrak{h},\mathfrak{h}_1,\ldots,\mathfrak{h}_M)$ operator (or $X$ for short) eliminates possible overcounting of the "shared" amplitudes. It enables to  re-express $\sigma_{\rm int}(\mathfrak{h})$ as
\begin{equation}
\sigma_{\rm int}(\mathfrak{h}) = \sigma_{\rm int}(\mathfrak{h}_i) + R(\mathfrak{h}_i) \;\;
(i=1,\ldots,M)\;,
\label{sisplit}
\end{equation}
where 
\begin{equation}
R(\mathfrak{h}_i) = 
^{(i)}\sum_{j=1}^M \; \sigma_{\rm int}(\mathfrak{h}_j) +
    X
\end{equation}
and $^{(i)}\sum_{j=1}^M$ designates the sum where the $i$-th element is neglected.
Consequently, we get 
\begin{equation}
    e^{\sigma_{\rm int}(\mathfrak{h})}|\Phi\rangle=
    e^{\sigma_{\rm int}(\mathfrak{h}_i)+R(\mathfrak{h}_i)}|\Phi\rangle \;\;(i=1,\ldots,M)\;.
    \label{ffqq1}
\end{equation}
Using Trotter formula we can approximate right hand side of (\ref{ffqq1}) for a given $j$  as 
\begin{equation}
     e^{\sigma_{\rm int}(\mathfrak{h})}|\Phi\rangle \simeq
     ( e^{R(\mathfrak{h}_i)/N}
     e^{\sigma_{\rm int}(\mathfrak{h}_i)/N})^N |\Phi\rangle \;.
\end{equation}
Introducing auxiliary operator $G^{(N)}_i$
\begin{equation}
    G^{(N)}_i=( e^{R(\mathfrak{h}_i)/N}
     e^{\sigma_{\rm int}(\mathfrak{h}_i)/N})^{N-1} e^{R(\mathfrak{h}_i)/N}
     \;\;(i=1,\ldots,M)\;,
\end{equation}
the "internal" wave function (\ref{ffqq1}) can be expressed as 
\begin{equation}
    e^{\sigma_{\rm int}(\mathfrak{h})}|\Phi\rangle \simeq G^{(N)}_i e^{\sigma_{\rm int}(\mathfrak{h}_i)/N} |\Phi\rangle
    \;\;(i=1,\ldots,M)\;.
    \label{ubi1}
\end{equation}
One should remember that  $G^{(N)}_i$ is a complicated function of all $\sigma_{\rm int}(\mathfrak{h}_j)\; 
(j=1,\ldots,M)$ and the above expression does not decouple $\sigma_{\rm int}(\mathfrak{h}_i)$ from the $G^{(N)}_i$ term. 
However, this expression may help define the practical way for determining computational blocks for flow equations. 
To see this, let us introduce expansion (\ref{ubi1}) to Eq.(\ref{duccstep2}) (with $H^{\rm eff}(\mathfrak{h})$ replaced by the $A$ operator), pre-multiply both sides by $\lbrack G^{(N)}_i \rbrack^{-1}$, and project onto $(P+Q_{\rm int}(\mathfrak{h}_i))$ sub-space, which leads to non-linear eigenvalue problems
\begin{widetext}
\begin{equation}
 (P+Q_{\rm int}(\mathfrak{h}_i)) \lbrack G^{(N)}_i\rbrack ^{-1} A G^{(N)}_i e^{\sigma_{\rm int}(\mathfrak{h}_i)/N} |\Phi\rangle \simeq E e^{\sigma_{\rm int}(\mathfrak{h})_i)/N} |\Phi\rangle
 \; (i=1,\ldots,M)
 \;.
 \label{upr1}
\end{equation}
\end{widetext}
We will utilize these equations as a computational blocks for the  DUCC flow. 
To make a practical use of Eqs. (\ref{upr1}) let us linearize them by defining the downfolded Hamiltonian $\Gamma_i^{(N)}$,
$\Gamma_i^{(N)}= (P+Q_{\rm int}(\mathfrak{h}_i)) \lbrack G^{(N)}_i\rbrack ^{-1} A G^{(N)}_i  (P+Q_{\rm int}(\mathfrak{h}_i))$ as a function of all
$\sigma_{\rm int}(\mathfrak{h}_j) \; (j=1,\ldots,M)$ from the previous flow cycle(s) ($pc$). We will symbolically designate this fact by using special notation for $\Gamma_i^{(N)}$ effective Hamiltonian, i.e., 
$\Gamma_{i}^{(N)}(pc)$ Hamiltonian. 
Now, we replace eigenvalue problems (\ref{upr1})) by optimization procedures described by  Eqs.(\ref{vqeduccx1}) which also offer an easy way to 
deal with "shared" amplitudes.  Namely, if in analogy to SR-CC sub-system flow we establish an ordering of
$\mathfrak{h}_i$ sub-algebras, with $\mathfrak{h}_1$ corresponding to the CAS closest to the wave function of interest, then 
in the $\mathfrak{h}_i$ problem we partition  (in analogy to Eq.(\ref{xcpamp})) set of parameters 
${\bm \theta}_N(\mathfrak{h}_i)$ into sub-set ${\bm \theta}_N^{\rm CP}(\mathfrak{h}_i)$ that refers to  common pool of  
amplitudes determined in preceding steps (say, for $\mathfrak{h}_j \; (j=1,\ldots,i-1)$) and sub-set  ${\bm \theta}_N^{\rm X}(\mathfrak{h}_i)$ that is uniquely determined in the $\mathfrak{h}_i$ minimization step, i.e, 
\begin{widetext}
\begin{equation}
\min_{{\bm \theta}_N^{\rm X}(\mathfrak{h}_i)}   
\langle\Psi({\bm \theta}_N^{\rm X}(\mathfrak{h}_i),{\bm \theta}_N^{\rm CP}(\mathfrak{h}_i))
|
\Gamma_{i}^{(N)}(pc)
|
\Psi({\bm \theta}_N^{\rm X}(\mathfrak{h}_i),{\bm \theta}_N^{\rm CP}(\mathfrak{h}_i))\rangle 
 \;\;(i=1,\ldots,M)\;,
\label{vqetrott}
\end{equation}
\end{widetext}
where $|\Psi({\bm \theta}_N^{\rm X}(\mathfrak{h}_i),{\bm \theta}_N^{\rm CP}(\mathfrak{h}_i))\rangle$ approximates 
$e^{\sigma_{\rm int}(\mathfrak{h}_i)/N} |\Phi\rangle$.
In this way, each computational block coupled into a flow corresponds to a minimization procedure that optimizes parameters
${\bm \theta}_N^{\rm X}(\mathfrak{h}_i)$ using quantum algorithms such as the VQE approach. 
At the end of the iterative cycle, once all amplitudes are converged, in contrast to the SR-CC flows, the energy is calculated using  $\mathfrak{h}_1$ problem
as an expectation value of the $\Gamma_{1}^{(N)}$ operator. 
The DUCC flow is  composed of classical computing steps where 
approximate second-quantized form of the 
$\Gamma_{i}^{(N)}(pc)$ operators (at the cost of additional similarity transformations or their approximate variants in small-size active space)  are calculated   and quantum computing steps, where cluster amplitudes are determined using the VQE algorithm. The discussed formalism introduces a broad class of control parameters, which define each computational step's dimensionality. These are the numbers of occupied/unoccupied active orbitals defining $\mathfrak{h}_i$ sub-algebras $x_{R_i}$/$y_{S_i}$, respectively. 
Similar results can be obtained by using the Zassenhaus formula. Moreover, the present formalisms, in analogy to the SR-CC flows,  can be extended to the time-domain.

An essential  feature of the DUCC flow equation is associated with the fact that each computational block (\ref{vqetrott})  can be encoded using a much smaller number of qubits compared to the full size of the global problem. 
In fact, the maximum size of the qubit register ($QR(\mathfrak{m}_{\rm max})$) required in DUCC quantum flow is associated with the maximum size of the sub-system and not with the size of the entire quantum system of interest
\begin{equation}
QR(\mathfrak{m}_{\rm max}) \ll QR(\mathfrak{g}^{(N)}) \;,
\label{qrsize}
\end{equation}
where $QR(\mathfrak{g}^{(N)})$ is the total number of qubits required to describe whole system.
This observation significantly simplifies the qubit  encoding of the effective Hamiltonians included in  quantum DUCC flows, especially in formulations based on the utilization of localized molecular basis set as discussed in Section II.C (for early quantum algorithms exploiting locality of interactions see Ref.\cite{ mcclean2014exploiting}).


\section{Conclusions}
This paper discussed the properties of SR-CC sub-system flow equations stemming from the SES-CC formalism for RHF reference functions. It was shown that flow equations define an alternative (to the canonical formulations) way of  introducing selected classes of higher-rank excitations based on system partitioning or choice of sub-system excitation sub-algebras corresponding to various active spaces.
An essential feature of the  SR-CC flow lies in the fact that flow equations can be built upon 
$\mathfrak{g}^{(N)}(x_R,y_S)$ sub-algebras with $x_R$ and $y_S$ chosen in a way that makes the flow tunable to available computational resources.  We also demonstrated that the idea of CC flow naturally extends to the time-domain, offering a possibility of performing calculations for the quantum system's time evolution affordably.
Interestingly, the ideas behind  SES-CC and SR-CC sub-system flows can also provide a deeper understanding of local CC formulations and the concept of "locality" of correlation effects. As explained in section II.B, the SR-CC flows based on the utilization of local molecular orbitals provide a rigorous way of defining sub-system through the effective Hamiltonian corresponding to $(i,j)$-determined SES, $\mathfrak{g}^{(N)}(2_{ij})$. The $(i,j)$-pair density matrix can be further used to calculate pair-natural orbitals and select leading excitation as postulated in the DLPNO-CCSD formulations.  We believe that the SR-CC flows defined by larger active spaces also provide a natural way of introducing higher-rank excitations, although maintaining linear-scaling of the resulting local CC formulation may not be possible. 
On the other hand, following the SR-CC flow philosophy for localized orbitals,
although numerically more expensive, may help in re-establishing the desired level of accuracy in perturbative (non-iterative) energy corrections  due to the higher-rank cluster excitations.

Due to the non-commutative nature of the general type unitary CC formulations, the direct extension of SR-CC sub-system flows to DUCC-type flow is a rather challenging endeavor. However,    utilizing Trotter formula in downfolding procedures lead to computationally feasible algorithms. 
In the quantum computing variant, the flow represents a sequence of coupled Hermitian eigenproblems, where diagonalization is replaced by the VQE-type optimization to obtain a corresponding  sub-set of amplitudes. In this formulation, the flow of quantum information corresponding to shared/external  amplitudes (defined for a given sub-system)
can be easily implemented at the level of the quantum circuit.
In analogy to the SR-CC flows, the DUCC-flows can be tuned to the available quantum resources. 
As such, the DUCC flows offer an interesting possibility of decomposing large-dimensionality problems
into a collection of reduced dimensionality computational blocks. 
We believe that the DUCC flow methods can significantly push the envelope of system-size tractable in quantum simulations. 
It should also be stressed that the SR-CC/DUCC flow methods allow one to enlarge the size of probed space systematically
while retaining the size-extensivity of the calculated in this way energies. This  feature is especially important in applications of DUCC flows to chemical reactions and extended systems. 

An exciting feature of the SES-CC formalism and CC flows is their universal character irrespective of the particular form of interactions defining correlated many-body systems. 

\section{acknowledgement}
This work was supported by the Quantum Science Center (QSC), a National Quantum Information Science Research Center of the U.S. Department of Energy (DOE).
Part of this work was supported by  the ``Embedding QC into Many-body Frameworks for Strongly Correlated  Molecular and Materials Systems'' project, which is funded by the U.S. Department of Energy, Office of Science, Office of Basic Energy Sciences (BES), the Division of Chemical Sciences, Geosciences, and Biosciences.

%
\appendix
\appendixpage
\section{ }   
In this Appendix we will analyze properties of the CC flow equations. Let us assume that flow equations involve $M$
eigenvalue problems defined by sub-algebras 
$\lbrace \mathfrak{h}_i \rbrace_{i=1}^{M}$
(see Eqs.({\ref{seseqh2})})
\begin{equation}
    H^{\rm eff}(\mathfrak{h}_i)
    e^{T_{\rm int}(\mathfrak{h}_i)}|\Phi\rangle = 
    E e^{T_{\rm int}(\mathfrak{h}_i)}|\Phi\rangle  \;\;
    (i=1,\ldots,M) \;.
\label{app0}
\end{equation}
The $i$-th computational block can be written in the following form 
\begin{equation}
(P+Q_{\rm int}(\mathfrak{h}_i)) 
\lbrack
e^{-T_{\rm ext}(\mathfrak{h}_i)}
H
e^{T_{\rm ext}(\mathfrak{h}_i)}-E
\rbrack
e^{T_{\rm int}(\mathfrak{h}_i)} |\Phi\rangle = 0 \;,
\label{app1}
\end{equation}
where the $T_{\rm ext}(\mathfrak{h}_i)$ operator in the above equation contains excitations 
from all remaining cluster operators $T_{\rm int}(\mathfrak{h}_j)\;\;(j=1,\ldots,M \; ;  j\neq i)$  not belonging to the set of excitations defining the $T_{\rm int}(\mathfrak{h}_i)$ operator.
After introducing the resolution of identity 
\begin{equation}
   e^{-T_{\rm int}(\mathfrak{h}_i)}
   e^{T_{\rm int}(\mathfrak{h}_i)}=1
\end{equation}
right to  the projection operator $(P+Q_{\rm int}(\mathfrak{h}_i))$ in Eq.(\ref{app1})
one obtains
\begin{equation}
(P+Q_{\rm int}(\mathfrak{h}_i)) 
e^{T_{\rm int}(\mathfrak{h}_i)}
\lbrack 
e^{-T}
H
e^{T} -E
\rbrack
|\Phi\rangle =0  \;,
\label{app2}
\end{equation}
where 
\begin{equation}
    T=T_{\rm int}(\mathfrak{h}_i)
    + T_{\rm ext}(\mathfrak{h}_i) \;.
    \label{app3}
\end{equation}
It should be stressed that the  $T$ operator is not carrying index of any sub-algebra, and $T$
is defined  as a sum of unique excitations defining 
$\lbrace T_{\rm int}(\mathfrak{h}_i)\rbrace_{i=1}^M$ operators.
We will symbolically represent $T$ as a union  of unique excitations originating in various $T_{\rm int}(\mathfrak{h}_i)$
\begin{equation}
    T= \bigcup_{i=1}^{M} T_{\rm int}(\mathfrak{h}_i)\;.
    \label{app3bb}
\end{equation}
This is a consequence of the CC flow definition. 
Given the fact that, the operator 
$e^{T_{\rm int}(\mathfrak{h}_i)}$ is non-singular and that 
\begin{equation}
(P+Q_{\rm int}(\mathfrak{h}_i)) 
e^{T_{\rm int}} = (P+Q_{\rm int}(\mathfrak{h}_i)) 
e^{T_{\rm int}}
(P+Q_{\rm int}(\mathfrak{h}_i)) \;,
\label{app4}
\end{equation}
the eigenvalue problem is equivalent {\it at the solution} to 
\begin{equation}
(P+Q_{\rm int}(\mathfrak{h}_i)) 
\lbrack 
e^{-T}
H
e^{T} -E
\rbrack
|\Phi\rangle =0
\label{app5}
\end{equation}
or in "standard" form of CC equations 
\begin{eqnarray}
(P+Q_{\rm int}(\mathfrak{h}_i)) 
\lbrack 
e^{-T}
H
e^{T}
\rbrack
|\Phi\rangle =0 && \label{app6} \\
E=\langle\Phi|e^{-T}
H
e^{T}|\Phi\rangle \;. \label{app7} 
\end{eqnarray}
Using Eqs.(\ref{app6}-\ref{app7}) one can draw the following conclusions:
\begin{itemize}
    \item {\bf Conclusion 1:} The CC flow equations (\ref{app0}) at the solution are equivalent to the standard connected CC equations for the $T$ cluster operator.
    \item {\bf Conclusion 2:} All ground-state eigenvalues of  computational blocks given by Eq.(\ref{app1}) are equal and at the solution assume the value of CC energy calculated using standard formulation for the $T$ cluster operator. 
\end{itemize}
The final remark concerns the size of the sub-space sampled by our flow equations, which is defined by excitations included in the $T$ cluster operator. For the CC flow defined by all possible $\mathfrak{g}^{(N)}(2_{R_i})$ sub-algebras, the 
$T$ operator contains all singles, doubles, and subsets of triple and quadruple exictations as shown in Eq.(\ref{safccsd2}). For the CC flow defined by all 
$\mathfrak{g}^{(N)}(3_{R_i})$ sub-algebras, the 
$T$ operator contains all singles, doubles, and triples, and subsets of quadruples, pentuples, and hextuples. The CC flows equations also offer a  flexibility in choosing sub-algebras involved in the flow, for example, the flow can involve various $\mathfrak{g}^{(N)}(x_{R_i},y_{S_i})$ sub-algebras. 
As long as $T$ is defined by unique excitations of the internal cluster operators corresponding to the sub-algebras involved in the flow,  conclusions 1 and 2 are still valid. 



\noindent \section{ }
The analysis in Appendix A can be extended to the CC flows in the  time domain, where the flow is composed of coupled time-dependent CC  equations
\begin{equation}
i\hbar \frac{\partial }{\partial t} e^{T_{\rm int}(\mathfrak{h}_i,t)} |\Phi\rangle = H^{\rm eff}(\mathfrak{h}_i,t)  e^{T_{\rm int}(\mathfrak{h}_i,t)} |\Phi\rangle \;,
(i=1,\ldots,M) \;.
\label{appb1}
\end{equation}
For a given sub-algebra $\mathfrak{h}_i$ time-dependent  equations can be cast in the form (we assume that spinorbitals are time-independent)
\begin{equation}
i\hbar (P+Q_{\rm int}(\mathfrak{h}_i))
\frac{\partial}{\partial t} e^{T_{\rm int}(\mathfrak{h}_i,t)} |\Phi\rangle =
H^{\rm eff}(\mathfrak{h}_i,t)  e^{T_{\rm int}(\mathfrak{h}_i,t)} |\Phi\rangle
\label{appb2}
\end{equation}
By  expanding $H^{\rm eff}(\mathfrak{h}_i,t)$,
Eq.(\ref{appb2}) can be rewritten as 
\begin{widetext}
\begin{equation}
(P+Q_{\rm int}(\mathfrak{h}_i))
e^{T_{\rm int}(\mathfrak{h}_i,t)} 
(P+Q_{\rm int}(\mathfrak{h}_i))
\lbrace
i\hbar \frac{\partial}{\partial t} T(t)
-e^{-T(t)} H e^{T(t)}
\rbrace
|\Phi\rangle
=0  \;,
\label{appb3}
\end{equation}
\end{widetext}
where $T(t)$ is defined as 
\begin{equation}
    T(t)= \bigcup_{i=1}^{M} T_{\rm int}(\mathfrak{h}_i,t)\;.
    \label{app3b}
\end{equation}
Using matrix representation of the $T_{\rm int}(\mathfrak{h}_i,t)$ operator in the $\mathfrak{h}_i$ generated CAS space denoted 
as $\bm{T}_{\rm int}(\mathfrak{h}_i,t)$ we get 
\begin{equation}
	{\rm det}
	(e^{\bm{T}_{\rm int}(\mathfrak{h}_i,t)})
	=e^{{\rm Tr}(\bm{T}_{\rm int}(\mathfrak{h}_i,t))} =1
	\label{appb4}
\end{equation}
for arbitrary time $t$. This is a consequence of the fact that $\bm{T}_{\rm int}(\mathfrak{h}_i,t)$ is a lower diagonal matrix with zeros on the diagonal.
Therefore,  Eq.(\ref{appb3}) is equivalent to standard time-dependent equations
\begin{equation}
(P+Q_{\rm int}(\mathfrak{h}_i))
\lbrace
i\hbar \frac{\partial}{\partial t} T(t)
-e^{-T(t)} H e^{T(t)}
\rbrace
|\Phi\rangle
=0  \;.
\label{appb3}
\end{equation}
Following the same reasoning as in Appendix A, we can state that:
\begin{itemize}
\item
{\bf Conclusion 3:}
The time-dependent CC flow approach (\ref{appb1}) is equivalent to standard representation of the time-dependent CC equations defined by the $T(t)$.
\end{itemize}. 

\noindent \section{ } 
In this appendix we discuss properties of the CC flows defined by approximate methods for solving computational blocks (\ref{app0}).
First, let us assume that the $T_{\rm int}(\mathfrak{h}_i)$ (defined by excitation level $m_i$) is approximated by operator $T_{\rm int}^{(A)}(\mathfrak{h}_i)$ which by  defined by excitations od maximum rank  $m(A)_i$ ($m(A)_i< m_i$).  
In this case, to obtain working equations for $T_{\rm int}^{(A)}(\mathfrak{h}_i)$ amplitudes, we will project Eq.(\ref{app1}) for a given $\mathfrak{h}_i$ onto 
$P+Q_{\rm int}^{(A)}(\mathfrak{h}_i)$, where $Q_{\rm int}^{(A)}(\mathfrak{h}_i)$ is a projection operator onto  excitations generated by $T_{\rm int}^{(A)}(\mathfrak{h}_i)$ when acting on $|\Phi\rangle$, i.e.,
\begin{equation}
(P+Q_{\rm int}^{(A)}(\mathfrak{h}_i))
\lbrack
H^{\rm eff}(\mathfrak{h}_i)
-E 
\rbrack
e^{T_{\rm int}^{\rm (A)}(\mathfrak{h}_i)} |\Phi\rangle = 0
\;,
\label{app8}
\end{equation}
where the full projection operator 
$Q_{\rm int}(\mathfrak{h}_i)$
can be partitioned as
\begin{equation}
  Q_{\rm int}(\mathfrak{h}_i)=  Q_{\rm int}^{(A)}(\mathfrak{h}_i)+R_{\rm int}^{(A)}(\mathfrak{h}_i) \;.
  \label{app9}
\end{equation}
Eq.(\ref{app8})
can be now re-written in the form 
\begin{widetext}
\begin{equation}
(P+Q_{\rm int}^{(A)}(\mathfrak{h}_i))
\lbrack
H^{\rm eff}(\mathfrak{h}_i)
-E 
\rbrack
(Q_{\rm int}^{(A)}(\mathfrak{h}_i)+R_{\rm int}^{(A)}(\mathfrak{h}_i))
e^{T_{\rm int}^{\rm (A)}(\mathfrak{h}_i)} |\Phi\rangle = 0
\;,
\label{app9}
\end{equation}
\end{widetext}
which clearly shows that truncating the full form of $T_{\rm int}(\mathfrak{h}_i)$
results  in equations that due to the presence of  
$R_{\rm int}^{(A)}(\mathfrak{h}_i)$-dependent term
are no longer representing eigenvalue problem.
Similar form of non-linear eigenvalue representation of CC has been analyzed by 
{\v Z}ivkovi\'c and Monkhorst in Ref.\cite{vzivkovic1978analytic}. 
Nevertheless, using similar reasoning as in the Appendix A, it can be shown that even in the case of standard approximations, Eq.(\ref{app8}) can be recast in the 
standard connected form
%
\begin{eqnarray}
(P+Q_{\rm int}^{\rm (A)}(\mathfrak{h}_i)) 
e^{-T^{\rm (A)}}
H
e^{T^{\rm (A)}} 
|\Phi\rangle =0 && \;, \label{capp6} \\
E=\langle\Phi|e^{-T^{\rm (A)}}
H
e^{T^{\rm (A)}}|\Phi\rangle \;, \label{capp7} 
\end{eqnarray}
where 
\begin{equation}
    T^{(A)}= \bigcup_{i=1}^{M} T_{\rm int}^{\rm (A)}(\mathfrak{h}_i)\;.
    \label{app3bb}
\end{equation}

\noindent \section{ } 

In this Appendix we  focus on  details of the numerical  realization  of the CC flow shown in Fig.\ref{fig2}.
Without a loss of generality,  we will focus on the "serial" flow shown in Fig.\ref{fig2}(a). The following steps are involved in the 
iterative process: 
\begin{enumerate}

\item {\bf Defining  sub-algebras/active spaces forming the flow:} In this step we define a set of active spaces corresponding to sub-algebras $\mathfrak{h}_i$ $i=1,\ldots,M$. 
In the flow  we solve for cluster operator $T$, $T= \bigcup_{i=1}^{M} T_{\rm int}(\mathfrak{h}_i)$.

\item {\bf Ordering of active spaces:} In serial flows,  an important step is associated with the establishing the importance of  active spaces. For example, this can be achieved using 
values of the MBPT(2) (second-order of many-body perturbation theory) correlation energy contributions in active spaces included in the flow. 

\item {\bf Initialization of the $T$ operator:} For this purpose we can use simple perturbative or low-rank CC approximations (CCSD).
 
\item {\bf Solving eigenvalue problems for active spaces:} In this step we solve for  $T_{\rm int}(\mathfrak{h}_i)$ update
by diagonalizing $e^{-T_{\rm ext}(\mathfrak{h}_i)}He^{T_{\rm ext}(\mathfrak{h}_i)}$ in the corresponding active space 
($T_{\rm ext}(\mathfrak{h}_i)=T-T_{\rm int}(\mathfrak{h}_i)$). In this step, resulting CI-type coefficients have to be transformed , using cluster 
analysis, to the $T_{\rm int}(\mathfrak{h}_i)$ amplitudes. 

\item {\bf Update of the global $T$ operator:} All $T_{\rm int}(\mathfrak{h}_i)$ $i=1,\ldots,M$ define a new one $T$ operator. 

\item{\bf Convergence check:} If $T$ operator satisfies convergence criteria, final value of correlation energy is calculated using either standard CC energy expression or diagonalizing any of the effective Hamiltonians involved in the flow. 
If does not, we repeat procedure from step (2). 

\end{enumerate}

The peak computational cost of the CC flow is defined by maximum-size active-space problem ($C_{\rm max}$). Therefore the cost per iteration is proportional to $M\times C_{\rm max}$. 
The CC flow equations approach offers a flexibility in the choice of the number of active spaces ($M$) and their size ($C_{\rm max}$) therefore providing a framework where the resulting computational model is  tuned to available computational resources.

It is also instructive to analyze the numerical cost of solving each computational block with approximate type methods such as CCSDT or CCSDTQ (see Appendix C). 
Without loss of generality, let us assume a CC flow defined by  $\mathfrak{g}^{(N)}(3_{R_i},y_{S_i})$ $i=1,\ldots,M$ problems.  We assume that for each $i=1,\ldots,M$ the number of active virtual orbitals is the same and equals $y$. 
For the CCSDT solver the upper-bound for the numerical cost of the flow, $FC({\rm CCSDT})$, is given by the formula, which is obtained by analizing the contributions from the  the most expensive terms:
\begin{equation}
FC({\rm CCSDT}) \le \alpha \times M \times {y \choose 3} \times  n_v^2 \;.
\label{apd1}
\end{equation}
Using the CCSDTQ solver for the same type of flow the numerical cost  upper-bound, $FC({\rm CCSDTQ})$, is given by the formula
\begin{equation}
FC({\rm CCSDTQ}) \le \beta \times M \times {y \choose 4} \times  n_v^2 \;.
\label{apd2}
\end{equation}
In (\ref{apd1}) and (\ref{apd2}), $\alpha$ and $\beta$ are constant pre-factors. 
Both upper-bounds depends on $M$ and $y$, whose values can be chosen to match available computational resources and/or provide the desired level of accuracy.



%

\end{document}